\documentclass[letterpaper]{article} 
\usepackage{aaai2027}  
\usepackage[hyphens]{url}  
\usepackage{graphicx} 
\usepackage{natbib}  
\usepackage{caption} 
\usepackage{algorithm}
\usepackage{algorithmic}

\usepackage{newfloat}
\usepackage{listings}
\DeclareCaptionStyle{ruled}{labelfont=normalfont,labelsep=colon,strut=off} 
\floatstyle{ruled}
\newfloat{listing}{tb}{lst}{}
\floatname{listing}{Listing}

\usepackage{booktabs}

\usepackage{hyperref}   

\usepackage{threeparttable}
\usepackage{amssymb}
\usepackage{amsmath}
\usepackage{booktabs}
\usepackage{multicol}
\usepackage{multirow}
\usepackage{graphicx}
\usepackage{tabularx}
\usepackage{amsthm}
\usepackage{subcaption}
\usepackage{siunitx}
\usepackage{array}
\usepackage{xcolor}
\usepackage{colortbl}

\title{M3: A State-Event Generative Foundation Model for Market Microstructure Dynamics}

\author{
    Yanzhi Zhang\textsuperscript{\rm 1,2,3},
    Yu Ma\textsuperscript{\rm 2},
    Yilin Cheng\textsuperscript{\rm 2},  
    Jian Li\textsuperscript{\rm 5},
    Yitong Duan\textsuperscript{\rm 2,4}\corresponding
}

\affiliations{
    \textsuperscript{\rm 1}Academy of Mathematics and Systems Science, Chinese Academy of Sciences\\
    \textsuperscript{\rm 2}Zhongguancun Academy\\
    \textsuperscript{\rm 3}University of Chinese Academy of Sciences\\
    \textsuperscript{\rm 4}Zhongguancun Institute of Artificial Intelligence\\
    \textsuperscript{\rm 5}IIIS, Tsinghua University\\
    zhangyanzhi20@mails.ucas.ac.cn, duanyitong@zgci.ac.cn
}

\makeatletter
\def\copyright@text{}
\makeatother

\begin{document}

\maketitle

\begin{abstract}
Market microstructure simulation aims to model how liquidity and order events coevolve in electronic financial markets. Since market data reveal only one realized trajectory, many important questions are inherently counterfactual and require realistic trajectory-level simulation. Existing financial generative models, however, often model order events and market states, such as the LOB, in isolation, overlooking the dynamic interaction between order events and liquidity in market microstructure. We propose the \textbf{M3} (\underline{M}arket \underline{M}icrostructure \underline{M}odel), a state-event generative foundation model for market microstructure dynamics. \textbf{M3} learns to generate future order-flow trajectories, while accounting for the evolving interaction between order events and limit order book. Trained on large-scale event level real stock market data, \textbf{M3} exhibits predictable scaling behavior, reproduces key market stylized facts, and enables practical simulation-based applications including forecasting, stress testing, and market-impact analysis. These results suggest a scalable foundation model paradigm for counterfactual market simulation at the microstructure level.

\begin{center}
\scriptsize
\begin{tabular}{@{}r@{\quad}l@{}}
\textbf{Code:} &
\href{https://github.com/ArthurZhang02/m3-market-microstructure}{
  \raisebox{-0.06em}{\includegraphics[height=0.28cm]{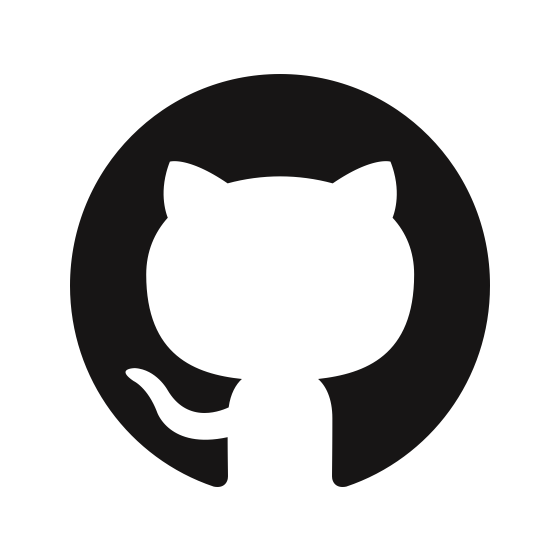}}
  \hspace{0.25em}\textbf{\texttt{ArthurZhang02/m3-market-microstructure}}
} \\
\textbf{Model:} &
\href{https://huggingface.co/Arthur210/M3-market-microstructure}{
  \raisebox{-0.06em}{\includegraphics[height=0.28cm]{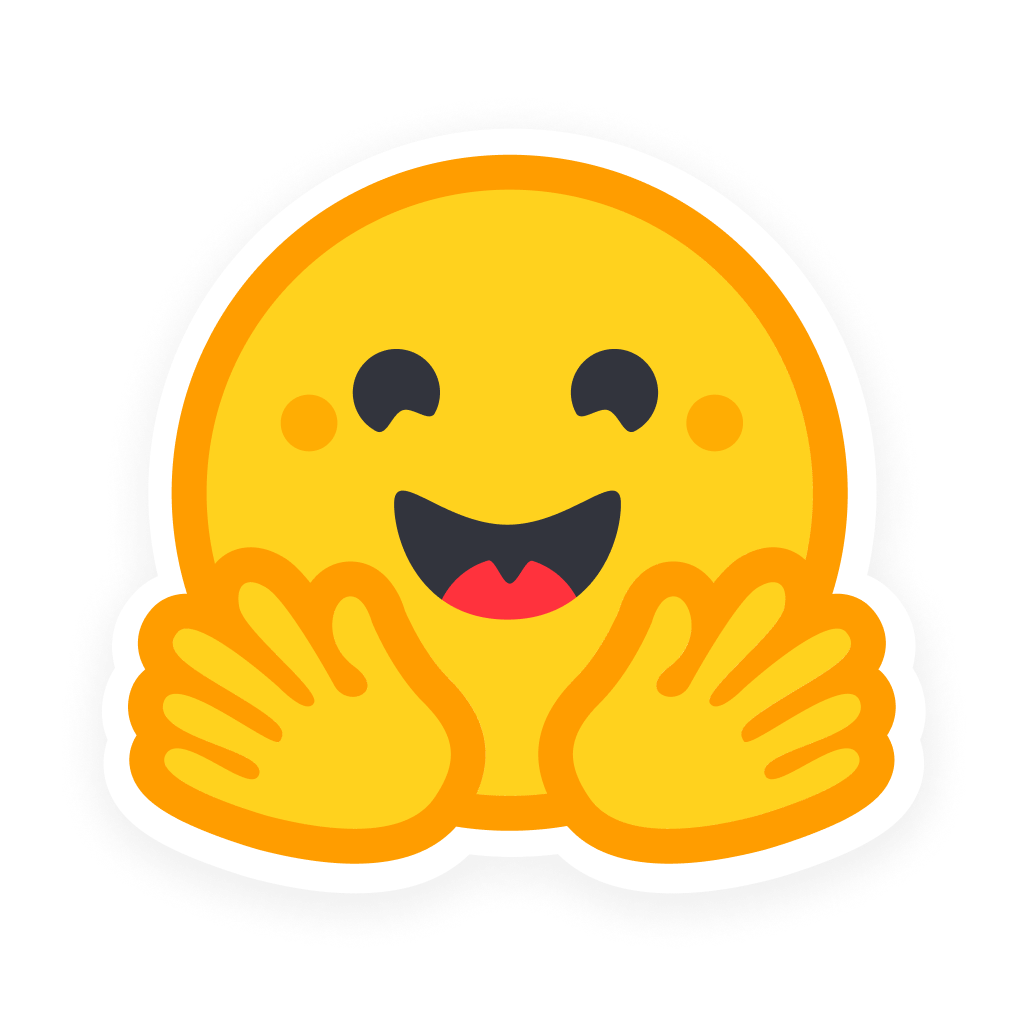}}
  \hspace{0.25em}\textbf{\texttt{Arthur210/M3-market-microstructure}}
}
\end{tabular}
\end{center}

\end{abstract}

\section{Introduction}

Financial markets are complex adaptive systems whose dynamics emerge from the interaction of heterogeneous participants, information acquisition, strategic trading, and liquidity conditions~\citep{grossman1980impossibility,kyle1985continuous,farmer1999frontiers,lo2004adaptive}. 
Historical market data only reveals one realized path. We do not know the range of alternative futures that could have unfolded, nor the market response to some interventions. Many practical questions are therefore inherently distributional or counterfactual: how might prices, spreads, and available liquidity evolve over the next interval; how would a large execution program reshape subsequent order flow and price impact; and how would a trading strategy perform across different market regimes?
Answering such questions requires trajectory-level market modeling: instead of predicting a single next outcome, the model should generate multiple market trajectories.


Modern electronic financial markets can be viewed as event-driven state-transition systems organized around the limit order book (LOB). Participants submit and cancel orders to express heterogeneous beliefs, manage inventory, and implement trading strategies. These events are processed by exchange matching rules and update the LOB, which records the liquidity available at different price levels. Conversely, the current LOB state---including depth, spread, and imbalance---shapes the execution outcome, price impact, and liquidity cost of subsequent orders.
At the microstructure level, order events supply, consume, and redistribute liquidity, while the LOB determines the execution outcome and price impact of incoming orders. 
Market dynamics therefore arise from a sequence of state transitions in which incoming orders update the book and the updated book state shapes subsequent order flow~\citep{gould2013limit,cont2013price,bouchaud2009markets}.
At the data level, this feedback loop is observed through two core modalities of market microstructure: order events and LOB states. Order events are irregular records involving price, volume, time, side, and action, whereas the LOB is a structured multi-level representation of liquidity. The heterogeneity of these two modalities, together with their tight coupling, makes market microstructure modeling challenging.

\begin{figure}[h]          
    \centering
    \includegraphics[width=0.97\columnwidth]{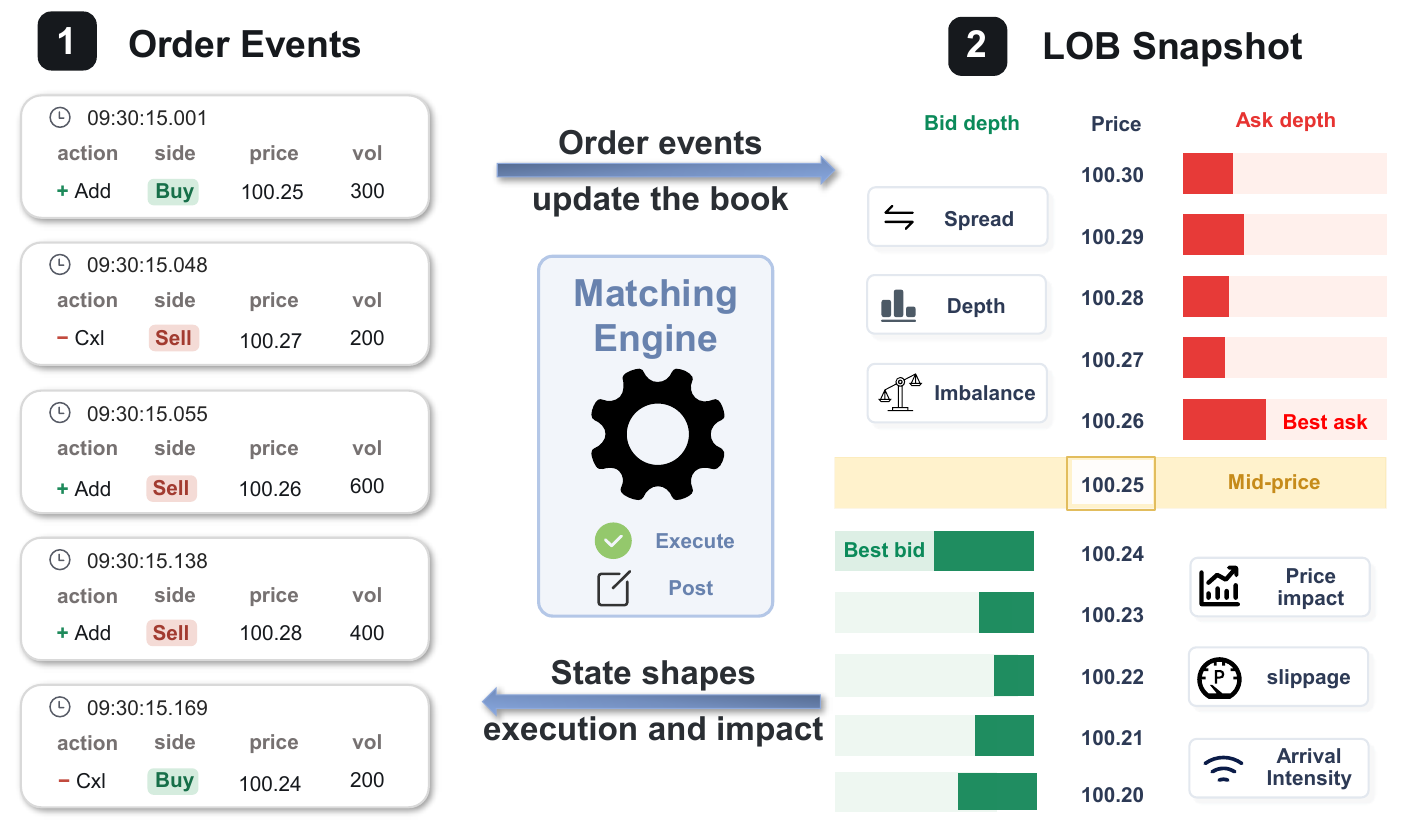}
    \caption{\textbf{Heterogeneous market microstructure data.}
    Unstructured order events are processed by matching engine to update the structured multi-level LOB, whose state in turn shapes execution outcomes and price impact.}
    \label{fig:hetr_data}
\end{figure}

Recent work has explored generative market simulation through
learned market agents, autoregressive order models, and diffusion-based
event generators~\citep{coletta2022worldagent,nagy2023generativeai,
zheng2024limitorderbookevent,li2025marsfinancialmarketsimulation,
kawawabeaudan2026tradefmgenerativefoundationmodel}.
These approaches motivate a market world-modeling perspective:
generating future order events and executing them under market rules
to study price dynamics, liquidity, and counterfactual interventions.

A central modeling choice is how to represent liquidity state
alongside an evolving order sequence. The LOB describes
the liquidity available before the sequence begins, while subsequent
orders determine how liquidity changes. We adopt a conditional
sequence formulation in which the initial state provides a fixed
context for autoregressive event generation within each window.

We propose \textbf{M3} (\underline{M}arket \underline{M}icrostructure
\underline{M}odel), a state-conditioned generative model of order sequences.
A vector-quantized tokenizer represents order events as discrete codes,
while an LOB encoder compresses a snapshot with 100 levels per side
into continuous state prefix tokens. A causal Transformer attends
to this prefix and the preceding order tokens to predict subsequent
events. The state prefix remains fixed within each generation window:
the model does not repeatedly re-encode the reconstructed book after
each generated event. This avoids feeding intermediate book
reconstruction errors back.

Generated tokens are decoded into orders and processed by a matching
engine to obtain book and price trajectories. For generation beyond
one window, the resulting book supplies the state prefix for the next
window. Thus, the fixed-prefix design applies within windows, while
state conditioning is refreshed at window boundaries. We use these
trajectories to evaluate return prediction, market dynamics, and
responses to order-flow interventions.

Our contributions are as follows:
\begin{itemize}
    \item \textbf{State-conditioned order-flow modeling.}
    We develop M3, combining vector-quantized order tokens with
    continuous prefixes derived from 100-level LOB snapshots.
    The model predicts order sequences conditioned on a fixed
    within-window liquidity context and preceding events.

    \item \textbf{Empirical scaling analysis.}
    We train M3 on approximately 31.9 billion order events and
    characterize the dependence of next-token prediction loss on
    model size and training-token budget using a Chinchilla-style fit.

    \item \textbf{Forecasting and simulation evaluation.}
    We evaluate M3 on prediction, stylized-fact reproduction,
    liquidity stress tests, and market-impact analysis. M3 demonstrates strong forecasting performance and market-simulation capabilities.
\end{itemize}

\begin{figure*}[ht]
    \centering
    \includegraphics[width=0.85\linewidth]{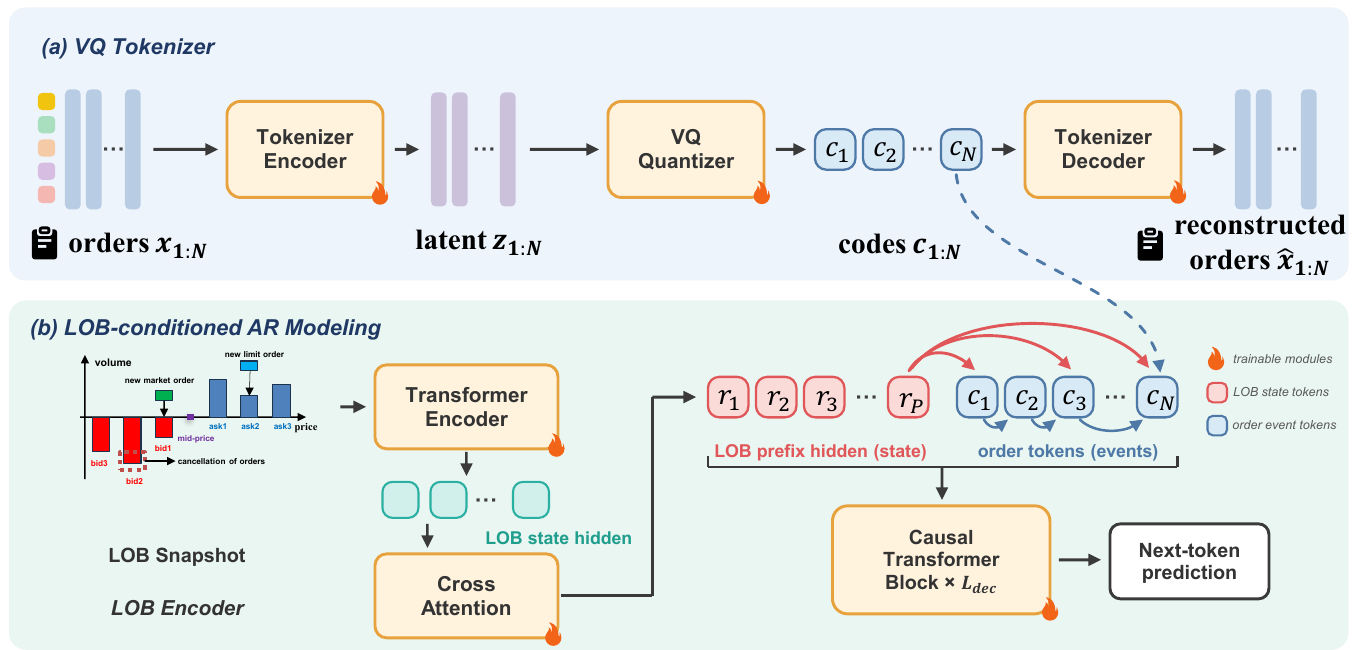}
\caption{\textbf{Architecture of Market Microstructure Model.}
(a) A VQ tokenizer discretizes raw order events into order tokens while preserving reconstructability through a decoder. 
(b) The autoregressive model conditions on the initial LOB state by encoding it into LOB prefix hidden, and then predicts subsequent order tokens.}
    \label{fig:main2222}
\end{figure*}

\section{Related Work}

\paragraph{Generative market models.}
Prior neural market simulators incorporate order history and liquidity information through conditional GANs~\citep{li2020stockgan}, state-dependent neural Hawkes processes~\citep{shi2022state}, and recurrent models of event attributes~\citep{hultin2023generative}. Other approaches learn generative agents within market simulators~\citep{coletta2022worldagent} or use diffusion to generate event types and arrival times~\citep{zheng2024limitorderbookevent}. More recently, TradeFM scales autoregressive event-stream modeling with a decoder-only Transformer and scale-invariant tokenization across equities~\citep{kawawabeaudan2026tradefmgenerativefoundationmodel}.

Neural market generators have long incorporated order history and liquidity state. Stock-GAN uses a conditional Wasserstein GAN with order-book constructions~\citep{li2020stockgan}. \citet{shi2022state} explicitly couple events and states in a state-dependent parallel neural Hawkes process. \citet{hultin2023generative} factor the distribution of each book transition into conditional distributions of event type, price level, size, and delay, modeled with recurrent networks. World-agent approaches learn market behavior within simulation environments~\citep{coletta2022worldagent}, while LOBDIF uses diffusion for event-type and arrival-time generation~\citep{zheng2024limitorderbookevent}.

The closest autoregressive precedents are \citet{nagy2023generativeai} and \citep{li2025marsfinancialmarketsimulation}. \citet{nagy2023generativeai} uses a deep state-space model for tokenized messages and book-state sequences and execute generated messages in JAX-LOB~\citep{nagy2023generativeai}. MarS's order model fuses an order embedding with a projection of antecedent ten-level LOB volumes and a midpoint embedding; generated orders are matched to obtain the subsequent book~\citep{li2025marsfinancialmarketsimulation}. 

\paragraph{Deep learning for limit order books.}
Deep learning has been widely applied to market microstructure modeling, particularly for short-horizon price prediction from limit order book (LOB) states. Representative models such as DeepLOB combine convolutional layers to capture the spatial structure of the LOB with recurrent modules to model temporal dependencies, achieving strong performance in mid-price movement prediction \citep{zhang2019deeplob}.

\paragraph{Discrete tokenization.}
Our model is related to discrete representation learning in generative modeling. VQ-VAE converts continuous observations into discrete codebook indices, allowing autoregressive models to learn priors over compact latent tokens instead of raw high-dimensional inputs \citep{oord2018neuraldiscreterepresentationlearning}. This idea has become a common design in multimodal generation: VQ-VAE-2, DALL-E, VQGAN, and video tokenizers represent images or videos as discrete visual tokens before sequence modeling \citep{razavi2019vqvae2,ramesh2021zeroshot,esser2021taming,yan2021videogpt,yu2023magvit}. More recently, Emu3 shows that images, videos, text, and actions can be unified as discrete token sequences and modeled with a single next-token prediction objective \citep{wang2026multimodal}. 

\paragraph{Agent-centric market simulation.}
Agent-centric models constitute a classical approach to financial market simulation, ranging from artificial stock markets with heterogeneous adaptive agents~\citep{arthur1997asset,lebaron1999timeseries} to microstructure models that simulate order submissions and cancellations through explicit or empirically calibrated rules~\citep{mike2008empirical}. Modern platforms such as ABIDES further provide high-fidelity discrete-event environments with realistic exchange protocols and latencies~\citep{byrd2019abideshighfidelitymarketsimulation}. 
Recent LLM-based simulators extend this paradigm by replacing hand-designed behavioral rules with generative agents that make trading decisions from market observations, news, policies, and fundamentals. ASFM combines LLM traders with an order-matching market~\citep{gao2024simulating}, while MarketSim scales this approach to large agent populations in a high-frequency continuous double-auction environment and separates strategic reasoning from execution~\citep{piao2026marketsim}.

\section{Preliminaries}

\subsection{Market Microstructure}


Modern electronic markets are typically organized around a \textbf{limit order book} (LOB), which records available buy and sell liquidity across multiple price levels. The highest bid and lowest ask define the \textbf{best bid} and \textbf{best ask}, respectively. Their difference is the \textbf{bid--ask spread}, while their average defines the \textbf{mid-price}. The volume available at each price level constitutes \textbf{market depth}. Together, spread and depth characterize market \textbf{liquidity}, or the ability to trade in size with limited price impact.
Schematic illustrations of the LOB structure and the order book feedback mechanism are provided in Figures~\ref{fig:hetr_data} and~\ref{fig:main2222}.

A central feature of electronic markets is the feedback loop between orders and the LOB states. Traders condition their actions on the current book, while order submissions and cancellations, together with the resulting executions, update available depth and the best quotes.
In turn, submitted orders, cancellations, and executions immediately reshape the book by consuming depth, moving the best quotes, or replenishing liquidity at specific levels. 
Market microstructure studies these fine-grained interactions and how they jointly govern price formation, liquidity provision, and information incorporation.







\subsection{Problem Setup}

\paragraph{Prediction objective.}
Given the LOB snapshot $L_0$ immediately before an observed sequence
of order events $\mathbf{o}_{1:n}$, our goal is to predict the joint distribution
of the next $H$ orders:
\begin{equation}
p_{\theta}(\mathbf{o}_{n+1:n+H}\mid L_0,\mathbf{o}_{1:n})
=
\prod_{j=1}^{H}
p_{\theta}(\mathbf{o}_{n+j}\mid L_0,\mathbf{o}_{1:n+j-1}).
\end{equation}
Here, $L_0$ describes the liquidity available before the observed
orders, which capture recent order-flow dynamics.

\paragraph{Problem Formulation.}
An order event is characterized by five basic attributes: action, side, price,
volume, and arrival time. We denote the n-th raw event arriving at time $\tau_n$ by
by ${\mathbf{o}}_n=\left[p_n,v_n,\tau_n,a_n,d_n\right]$, where \(p_n\), \(v_n\), and \(\tau_n\) are the raw order price, raw order size,
and submission time, respectively. And $a_n \in \{0, 1\}$ denotes the order action ($0$: cancel, $1$: add), and $d_n \in \{0, 1\}$ denotes the side ($0$: ask, $1$: bid).

We do not distinguish between limit and market orders; all submitted events are represented within a unified event space as feature vectors, denoted as 
$\mathbf{x}_n=[\rho_n^{\mathrm{open}},\ell_n^v,\Delta\tau_n,a_n,d_n].$. However, departing from the models \citep{kawawabeaudan2026tradefmgenerativefoundationmodel,li2025marsfinancialmarketsimulation} that normalize prices against the prevailing mid-price, we use the \textbf{daily open price} $m^{\mathrm{open}}$ as our anchor. Accordingly, the relative price feature is defined as $\rho^{\mathrm{open}}_n = (p_n-m^{\mathrm{open}})/m^{\mathrm{open}}$. For the remaining features, order volume is log-transformed as $\ell^v_n = \log(1+v_n)$, and the inter-arrival time is given by $\Delta \tau_n = \tau_n-\tau_{n-1}$.

The LOB snapshot at the beginning of window $k$ is represented by
an tensor $L_k\in\mathbb{R}^{2\times d_L\times2}$.
For side $b\in\{\mathrm{ask},\mathrm{bid}\}$ and level
$\ell\in\{1,\ldots,d_L\}$, its price and volume channels are
$(L_k)_{b,\ell,1}=p^{(k)}_{b,\ell},$ $ (L_k)_{b,\ell,2}=v^{(k)}_{b,\ell}.$
We use $d_L=100$ levels per side
in all experiments, giving $2d_L=200$ book positions.



\section{Methods}

\subsection{Overview}

Our framework generates executable order trajectories conditioned on an initial market state. It consists of two main components: a vector-quantized order tokenizer and an auto-regressive model with an LOB encoder. The tokenizer maps continuous order events into discrete tokens and decodes generated tokens back into order events. In parallel, an LOB encoder converts the initial order-book snapshot \(L_0\) and auxiliary state variables \(\mathbf{s}_0\) into a sequence of continuous prefix embeddings. A causal Transformer then models the conditional distribution of future order tokens. During inference, generated tokens are decoded and legalized into executable orders and replayed through the matching engine, producing both an order trajectory and its corresponding LOB states.

\subsection{Tokenization}

\paragraph{Order tokenization with vector quantization.}
A natural approach to order-flow tokenization is to discretize price, volume, time, action type, and side separately and combine their bin indices into a joint order token~\citep{li2025marsfinancialmarketsimulation,kawawabeaudan2026tradefmgenerativefoundationmodel}. However, increasing the resolution of multiple attributes causes the joint vocabulary to grow multiplicatively, making fine-grained reconstruction inefficient.

More importantly, mid-relative price discretization introduces a \textbf{training--inference mismatch}. During \textbf{training}, each relative price is computed using the ground-truth mid-price. During \textbf{inference}, the mid-price must instead be recovered by replaying previously generated and reconstructed orders through the simulated book. Reconstruction errors in these orders can perturb the simulated LOB and shift its mid-price, causing subsequent prices to be decoded using an inaccurate anchor and creating a feedback loop of accumulating errors. Moreover, continuously recentering prices around the current mid-price \textbf{removes information about their global intraday position}: the same relative-price token may correspond to substantially different absolute market levels at different stages of the trajectory, making it harder for the model to capture cumulative price movements and long-range market context.

We therefore normalize prices relative to the daily open price, which is fixed and identically available during \textbf{training} and \textbf{inference}. This avoids dependence on the reconstructed mid-price and preserves cumulative price movements from the open. However, the open-relative price distribution is substantially wider, so uniform binning would require either many bins or coarse resolution. We instead use vector quantization to learn a compact, non-uniform codebook over the joint order-feature space. As shown in the appendix, \textbf{the simulated-mid bin baseline suffers substantially larger reconstruction errors than its oracle counterpart, while the proposed open-anchor VQ tokenizer achieves higher reconstruction fidelity.}



We now describe the construction of the order tokenizer.
Each order event is represented by a normalized feature vector
combining continuous and categorical attributes. Given an order-flow
window
$
X=(\mathbf{x}_1,\ldots,\mathbf{x}_N),
$
the tokenizer maps each event $\mathbf{x}_n$ to a discrete token
$c_n\in\{0,\ldots,K-1\}$, where $K$ is the codebook size.
The first three entries of $\mathbf{x}_n$, namely
$\rho_n^{\mathrm{open}}$, $\ell_n^v$, and $\Delta\tau_n$,
are continuous, while $a_n$ and $d_n$ are categorical.
Following~\citet{shi2025kronosfoundationmodellanguage},
tokenization is performed by a causal encoder
$E_{\theta_{\mathrm{enc}}}$, a vector quantizer $Q$,
and a causal decoder $D_{\phi}$.
The encoder produces contextual latents:
$
\mathbf{z}_n
=
E_{\theta_{\mathrm{enc}}}(\mathbf{x}_{\leq n})
\in\mathbb{R}^{d_z}.
$
The quantizer maintains a codebook
$
\mathcal{E}
=
\{\mathbf{e}_k\}_{k=0}^{K-1},
\qquad
\mathbf{e}_k\in\mathbb{R}^{d_z},
$
and assigns each latent to its nearest entry:
$
c_n
=
\arg\min_{k\in\{0,\ldots,K-1\}}
\left\|\mathbf{z}_n-\mathbf{e}_k\right\|_2^2,
\quad
\mathbf{q}_n=\mathbf{e}_{c_n}.
$
Here, $c_n$ is the discrete order token and $\mathbf{q}_n$
is its quantized embedding. The causal decoder reconstructs
the input features as
$
\widehat{\mathbf{x}}_n
=
D_{\phi}(\mathbf{q}_{\leq n}).
$
Further details on the codebook size, exponential moving average
(EMA) updates, and codebook utilization are provided in the appendix.


\begin{figure*}[h]
  \centering
  \includegraphics[width=0.90\linewidth]{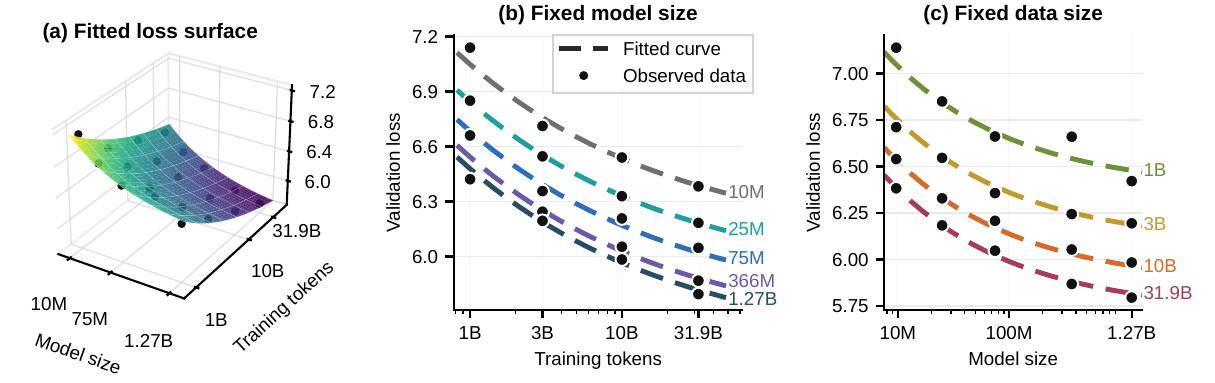}
\caption{\textbf{Chinchilla-style scaling law.} Fitted validation-loss surface over model size and training tokens. Black dots denote observed runs, while dashed curves show fitted scaling-law slices at fixed model sizes and fixed dataset sizes.}
  \label{fig:chinchilla_scaling_law}
\end{figure*}

\subsection{Autoregressive Model Structure}

After tokenization, each order sequences window of $N$ events is represented as a sequence of
discrete VQ tokens
$
C=(c_1,\ldots,c_N),\quad c_n\in\{0,\ldots,K-1\}.
$

\paragraph{LOB-conditioned auto-regressive model.}

To model how the prevailing liquidity state shapes subsequent order-flow dynamics, we introduce an LOB-conditioned auto-regressive model. We encode the market state into continuous prefix tokens that share the same causal Transformer backbone with generated order tokens. This enables the model to learn how the current liquidity configuration influences future order events, while generated orders subsequently drive LOB evolution through the matching engine.

For LOB snapshot $L_k$, the LOB encoder constructs a feature vector
for each side--level pair: 
\begin{equation}
\mathbf{u}^{(k)}_{b,\ell}
=
\begin{bmatrix}
\operatorname{clip}\!\left(
\dfrac{p^{(k)}_{b,\ell}-m_k}{m_k},-0.20,0.20\right)\\
\log\!\left(1+v^{(k)}_{b,\ell}\right)\\
\end{bmatrix}
\in\mathbb{R}^{2},
\end{equation}
where $m_k$ is the reference mid-price associated with $L_k$.
The initial window uses $k=0$.


Each feature vector is projected to the autoregressive hidden
dimension $d_{\mathrm{AR}}$ and augmented with side embeddings,
level-index embeddings, and projected auxiliary state features
$\mathbf{s}_k$. A Transformer encoder models interactions
among the $2d_L$ book positions. Learned queries then pool these
positions into $P$ continuous prefix embeddings:
$
\mathbf{r}^{\mathrm{lob}}_{k,1:P}
=G_{\psi}(L_k,\mathbf{s}_k)
\in\mathbb{R}^{P\times d_{\mathrm{AR}}}.
$

The LOB prefix embeddings precede the BOS and order token embeddings. Let $c_0=c_{\mathrm{BOS}}$ denote the start token.
Training minimizes the mean order token negative log-likelihood:
\begin{equation}
\mathcal{L}_{\mathrm{AR}}
=-\frac{1}{N}\sum_{n=1}^{N}\log p_{\theta_{\mathrm{AR}}}
\left(
c_n
\mid
\mathbf{r}^{\mathrm{lob}}_{0,1:P},c_0,c_{<n}
\right).
\end{equation}
During training, labels at the LOB-prefix and \(\mathrm{BOS}\) positions are masked.




\paragraph{Matching engine.}
We implement a matching mechanism capable of processing continuous order streams during both the call-auction
and continuous-trading phases.
The engine legalizes decoder outputs and updates persistent, full-depth
bid and ask books using price priority and anonymous FIFO queues.
Appendix details the initialization, legalization, and matching rules.

\section{Experiments}

\subsection{Dataset}
We use order-level market data for 800 non-overlapping CSI 300 and CSI 500 equities. M3 is trained on approximately 31.9 billion order events from January 2024 to November 2025, and evaluated on the held-out December 2025  and January--February 2026 periods. Details of the dataset and preprocessing procedure are provided in the appendix. Evaluation periods and sample coverage are specified for each experiment.




\subsection{Scaling Law}


Prior studies on LLM scaling have shown that language modeling loss follows predictable trends as a function of model size, data scale, and training compute~\citep{kaplan2020scalinglawsneurallanguage,hoffmann2022trainingcomputeoptimallargelanguage}. 
To examine whether similar scaling behavior emerges in the \textbf{financial domain}, we fit a Chinchilla's scaling law 
$
\mathcal{L}_{\mathrm{val}}(M,D)=E+AM^{-\alpha}+BD^{-\beta},$ 
to the validation losses observed for our market models. 
The resulting fitted loss function is 
\begin{equation}
\label{eq:scaling}
\mathcal L_{\mathrm{val}}(M, D) = 5.3697 + \frac{1.5987}{M^{0.3795}} + \frac{1.0027}{D^{0.3117}},
\end{equation}
where $M$ denotes the number of model parameters in millions and $D$ denotes the number of training tokens in billions. Figure~\ref{fig:chinchilla_scaling_law} visualizes the predictions implied by the fitted scaling law. 

To assess predictive validity,
we performed leave-one-model-size-out validation and a residual bootstrap over the fitted parameters. Refitting Eq.~\eqref{eq:scaling} on each four-size subset predicts the held-out size with a mean
absolute relative error of $0.66\%$ across folds; a fit restricted to models up to 366M predicts the held-out
1.27B within $0.53\%$. 

\paragraph{Predictability across token positions.}

\begin{figure}[t]
    \centering
    \includegraphics[width=0.90\columnwidth]
    {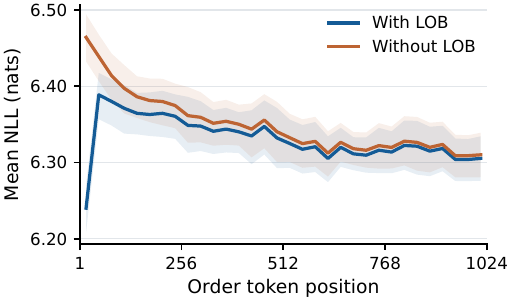}
    \caption{
    \textbf{Predictability across token positions.}
    Mean teacher-forcing NLL of model in non-overlapping 32-token bins,
    using validation windows with or without the LOB prefix. Shading denotes 95\% confidence intervals.    
    }
    \label{fig:position_nll}
\end{figure}

Figure~\ref{fig:position_nll} shows the teacher-forced NLL of AR model
across order windows. With LOB conditioning, NLL starts low, rises
rapidly, and then declines gradually. Removing the LOB prefix eliminates
the initial dip, and the gap narrows as the order history grows.
This pattern is consistent with book information being most useful
for predicting nearby events.

\subsection{Stylized Facts}

We assess whether generated order flow reproduces distributional and
temporal regularities of financial markets, including heavy tails,
aggregation effects, weak linear return dependence, and volatility
clustering~\citep{mandelbrot1963variation,cont2001empirical,ding1993long}.
We replay generated orders through the matching engine and compare the
induced mid-price paths with observed market paths. For horizon $\Delta$,
the mid-price log return is
$r_{t,\Delta}=\log(m_{t+\Delta}/m_t)$.
All M3 results in this section use the Large model.

\begin{figure}[t]
    \centering
    \includegraphics[width=0.98\linewidth]
    {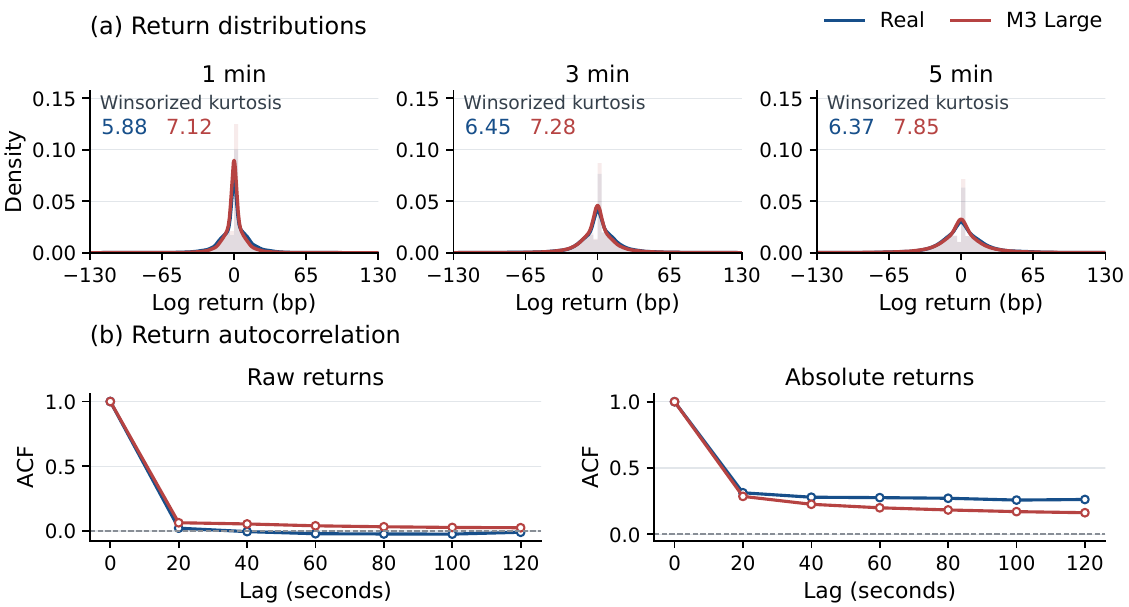}
\caption{
\textbf{Stylized facts of generated market trajectories.}
(a) Mid-price log-return distributions at 1, 3, and 5 minutes
for M3 Large and market paths. 
(b) Autocorrelations of 20-second raw and absolute log returns. 
}
\label{fig:stylized_facts}
\end{figure}

Figure~\ref{fig:stylized_facts}(a) shows sharply peaked, heavy-tailed
returns at 1, 3, and 5 minutes. Annotations report Pearson kurtosis after
0.5\% winsorization in each tail. Return dispersion increases with the
aggregation horizon. 
M3's winsorized kurtosis remains above the Gaussian value of three and increases over this range.
In Figure~\ref{fig:stylized_facts}(b), raw-return correlations are small
beyond lag zero, whereas absolute-return correlations remain positive,
consistent with weak linear dependence and volatility clustering.

Table~\ref{tab:tradefm_fidelity_with_kurtosis} compares M3 with
Zero-Intelligence (ZI)~\citep{2005farmerzi,GodeSunder1993zi} and Compound Hawkes (Hawkes)~\citep{swishchuk2020compound,bacry2015hawkes} baselines. In Panel A, M3 achieves the lowest Kolmogorov--Smirnov
and Wasserstein distances at 10 and 20 seconds, while ZI performs best at
120 seconds. 
Panel B reports raw and winsorized kurtosis.
At 5 minutes, M3 is closest in raw kurtosis, while
Hawkes is closest after winsorization.

\begin{table}[t]
\centering
\small
\setlength{\tabcolsep}{3pt}
\renewcommand{\arraystretch}{1.06}

\begin{tabular*}{\columnwidth}{@{\extracolsep{\fill}}lrrrrrr@{}}
\toprule
\multicolumn{7}{@{}l}{\textbf{Panel A: Log-return distributional fidelity}} \\
\addlinespace[1pt]
& \multicolumn{3}{c}{KS $\downarrow$}
& \multicolumn{3}{c}{$W_1$ (bp) $\downarrow$} \\
\cmidrule(lr){2-4}\cmidrule(lr){5-7}
$\Delta t$ (s) & ZI & Hawkes & M3 & ZI & Hawkes & M3 \\
\midrule
10 & 0.175 & \underline{0.149} & \textbf{0.122} & 3.03 & \underline{2.66} & \textbf{1.50} \\
20 & 0.178 & \underline{0.168} & \textbf{0.148} & \underline{3.54} & 3.67 & \textbf{2.35} \\
60 & \textbf{0.206} & 0.220 & \underline{0.212} & \underline{5.33} & 6.89 & \textbf{5.18} \\
120 & \textbf{0.264} & 0.291 & \underline{0.289} & \textbf{8.16} & 11.32 & \underline{8.96} \\
\end{tabular*}

\vspace{2pt}

\begin{tabular*}{\columnwidth}{@{\extracolsep{\fill}}rlrrrr@{}}
\midrule
\multicolumn{6}{@{}l}{\textbf{Panel B: Aggregational Gaussianity}} \\
\addlinespace[1pt]
$\Delta t$ (min) & Statistic & Real & ZI & Hawkes & M3 \\
\midrule
3 & Kurtosis & 17.48 & \textbf{11.57} & 74.96 & \underline{35.13} \\
 & Winsorized & 6.45 & \textbf{5.96} & 7.69 & \underline{7.28} \\
\addlinespace[1pt]
5 & Kurtosis & 28.69 & \underline{10.01} & 67.18 & \textbf{23.06} \\
 & Winsorized & 6.37 & \underline{5.45} & \textbf{7.02} & 7.85 \\
\bottomrule
\end{tabular*}

\par\vspace{2pt}
\parbox{\columnwidth}{\scriptsize\raggedright
M3: Large. \textbf{Bold}/\underline{underline}: best/second best.
Panel B ranks closeness to Real; winsorization clips 0.5\% per tail.
}
\caption{Log-return distributional fidelity and aggregational Gaussianity.}
\label{tab:tradefm_fidelity_with_kurtosis}
\end{table}





\begin{table*}[t]
\centering
\small
\setlength{\tabcolsep}{4pt}
\renewcommand{\arraystretch}{1.10}
\definecolor{PredictionBand}{HTML}{F0F2F5}
\definecolor{PredictionOurs}{HTML}{E7EDF6}
\begin{tabular}{p{0.40\textwidth}*{6}{>{\raggedleft\arraybackslash}p{\dimexpr(\textwidth-0.40\textwidth-14\tabcolsep)/6\relax}}}
\toprule
& \multicolumn{2}{c}{\textbf{1 minute}}
& \multicolumn{2}{c}{\textbf{2 minutes}}
& \multicolumn{2}{c}{\textbf{3 minutes}} \\
\cmidrule(lr){2-3}\cmidrule(lr){4-5}\cmidrule(lr){6-7}
Model or signal & IC & RankIC & IC & RankIC & IC & RankIC \\
\midrule
\rowcolor{PredictionBand}\multicolumn{7}{l}{\footnotesize\bfseries Zero-shot time-series foundation models} \\
\addlinespace[2pt]
Chronos-2$^{\dagger}$~\citep{ansari2025chronos2} & $0.0464$ & $0.0033$ & $0.0409$ & $0.0458$ & $0.0418$ & $0.0735$ \\
TimesFM-2.5$^{\dagger}$~\citep{google2025timesfm25} & $-0.0555$ & $-0.0269$ & $0.0029$ & $0.0193$ & $-0.0104$ & $0.0710$ \\
Toto-1.0~\citep{cohen2025toto} & $-0.0028$ & $0.0168$ & $-0.0010$ & $-0.0004$ & $0.0010$ & $0.0033$ \\
\addlinespace[3pt]
\rowcolor{PredictionBand}\multicolumn{7}{l}{\footnotesize\bfseries Fine-tuned time-series foundation models} \\
\addlinespace[2pt]
Chronos-2$^{\dagger}$ & $0.0850$ & $0.0562$ & $0.0246$ & $0.0291$ & $-0.0014$ & $0.0290$ \\
TimesFM-2.5$^{\dagger}$ & $0.0513$ & $0.0199$ & $0.0292$ & $0.0225$ & $0.0166$ & $0.0230$ \\
Toto-1.0 & $-0.0206$ & $-0.0201$ & $0.0309$ & $0.0334$ & $0.0639$ & $0.0574$ \\
\addlinespace[3pt]
\rowcolor{PredictionBand}\multicolumn{7}{l}{\footnotesize\bfseries Financial foundation model} \\
\addlinespace[2pt]
Kronos-base~\citep{shi2025kronosfoundationmodellanguage} & $-0.0731$ & $-0.0326$ & $0.0207$ & $0.0356$ & $0.0454$ & $0.0714$ \\
\addlinespace[3pt]
\rowcolor{PredictionBand}\multicolumn{7}{l}{\footnotesize\bfseries Microstructure signals} \\
\addlinespace[2pt]
Weighted-mid displacement & $0.0741$ & $0.0608$ & $\underline{0.0677}$ & $0.0746$ & $\underline{0.0656}$ & $0.0750$ \\
Depth-normalized OFI (L1) & $0.0736$ & $0.0947$ & $-0.0139$ & $0.0166$ & $-0.0478$ & $-0.0255$ \\
Queue imbalance (L1) & $0.0315$ & $0.0491$ & $0.0489$ & $0.0655$ & $0.0484$ & $0.0681$ \\
\addlinespace[3pt]
\rowcolor{PredictionBand}\multicolumn{7}{l}{\footnotesize\bfseries Order foundation models} \\
\addlinespace[2pt]
\rowcolor{PredictionOurs}
\textbf{M3 Tiny (ours)} & $\underline{0.1792}$ & $\underline{0.1297}$ & $0.0596$ & $\underline{0.1149}$ & $0.0521$ & $\underline{0.0905}$ \\
\rowcolor{PredictionOurs}
\textbf{M3 Large (ours)} & $\mathbf{0.2486}$ & $\mathbf{0.1790}$ & $\mathbf{0.1068}$ & $\mathbf{0.1661}$ & $\mathbf{0.1262}$ & $\mathbf{0.1351}$ \\
\addlinespace[3pt]
\bottomrule
\end{tabular}
\par\vspace{3pt}
\parbox{\textwidth}{\scriptsize\raggedright
The highest and second-highest
values in each column are \textbf{bold} and \underline{underlined},
respectively.\\
Model forecasts use MC32 price means. $\dagger$ indicates approximate
quantile sampling.
}
\caption{Return prediction by M3, foundation-model baselines, and
microstructure signals on the December 2025 panel.}
\label{tab:return_prediction}
\end{table*}

\begin{figure}[t]
    \centering
    \includegraphics[width=0.98\linewidth]
    {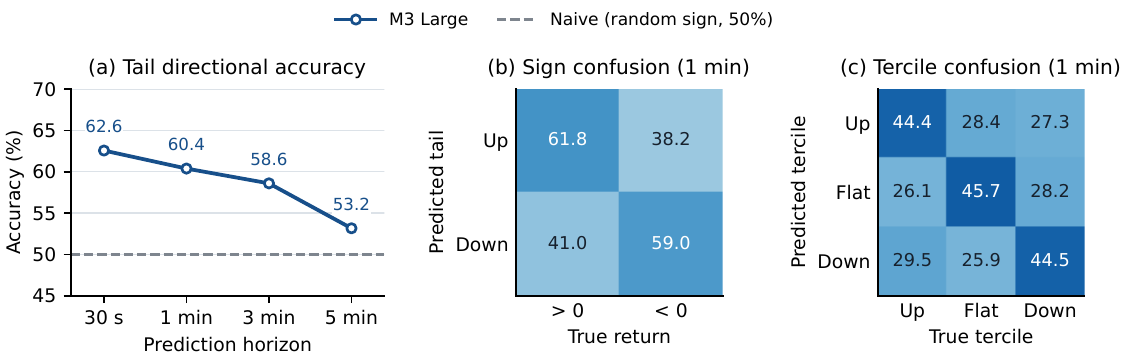}
    \caption{
    \textbf{Mid-price predictive diagnostics.}
    (a) Tail directional accuracy across horizons for M3 Large.
    (b) Sign confusion matrix at 1-min horizon.
    (c) Return-tercile confusion matrix at 1-min horizon.
    }
    \label{fig:midprice_prediction}
\end{figure}

\subsection{Prediction}

Prediction in financial markets requires modeling both the direction and the uncertainty of future prices. M3 predicts future prices by generating order trajectories and replaying them through the matching engine. The resulting mid-price paths provide forecasts of price movements, returns, and realized volatility~\citep{li2025marsfinancialmarketsimulation}.


\begin{figure}[t!]
    \centering
    \includegraphics[width=0.95\linewidth]{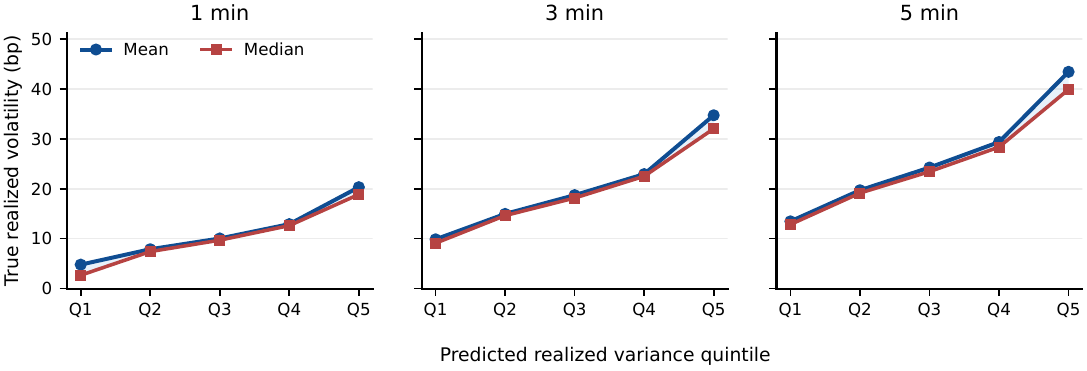}
\caption{ \textbf{Predicted realized volatility.} Mean and median realized volatility of the future paths, grouped by predicted realized-variance quintile. 
}
\label{fig:volatility}
\end{figure}

\paragraph{Mid-price prediction.}
We average 32 generated prices per prompt (MC32) to obtain an ensemble
return forecast. Figure~\ref{fig:midprice_prediction}(a) evaluates
directional accuracy in the upper and lower forecast terciles,
conditional on nonzero realized returns, against a 50\% random-sign
reference. Figure~\ref{fig:midprice_prediction}(b,c) report sign and return-tercile confusion matrices at one minute. 
The tail-accuracy definition is given in Appendix~\ref{app:midprice_prediction}.

\paragraph{Return prediction.}
We measure return-prediction quality across both stocks and forecast
origins. Let $\mathcal I$ index stocks and $\mathcal T$ index forecast
origins, each specified by a date and time. At horizon $h$, let
$\mathcal P_h\subseteq\mathcal I\times\mathcal T$ contain the valid
stock--origin pairs. The panel IC is
\begin{equation}
\mathrm{IC}_{\mathrm{panel}}(h)
=\operatorname{Corr}_{(i,t)\in\mathcal P_h}
\!\left(\widehat R_{i,t,h},R_{i,t,h}\right),
\label{eq:panel_ic}
\end{equation}
where $\operatorname{Corr}$ denotes Pearson or Spearman correlation and
$\widehat R_{i,t,h}$ and $R_{i,t,h}$ are predicted and realized
returns.
Both IC and RankIC are computed once over the whole panel.

Table~\ref{tab:return_prediction} compares M3 Tiny and Large with
general time-series foundation models, financial foundation model Kronos and microstructure signals. M3 conditions on the LOB and order flow; the baselines
use historical midpoint bars through their native interfaces.
The general models are evaluated before and after fine-tuning,
while Kronos retains its pretrained weights. All point forecasts use
MC32 price means. M3 Large achieves the highest IC and RankIC at all reported horizons.

\begin{figure*}[t!]
\centering
\begin{subfigure}{0.36\linewidth}
    \centering
    \includegraphics[width=\linewidth]{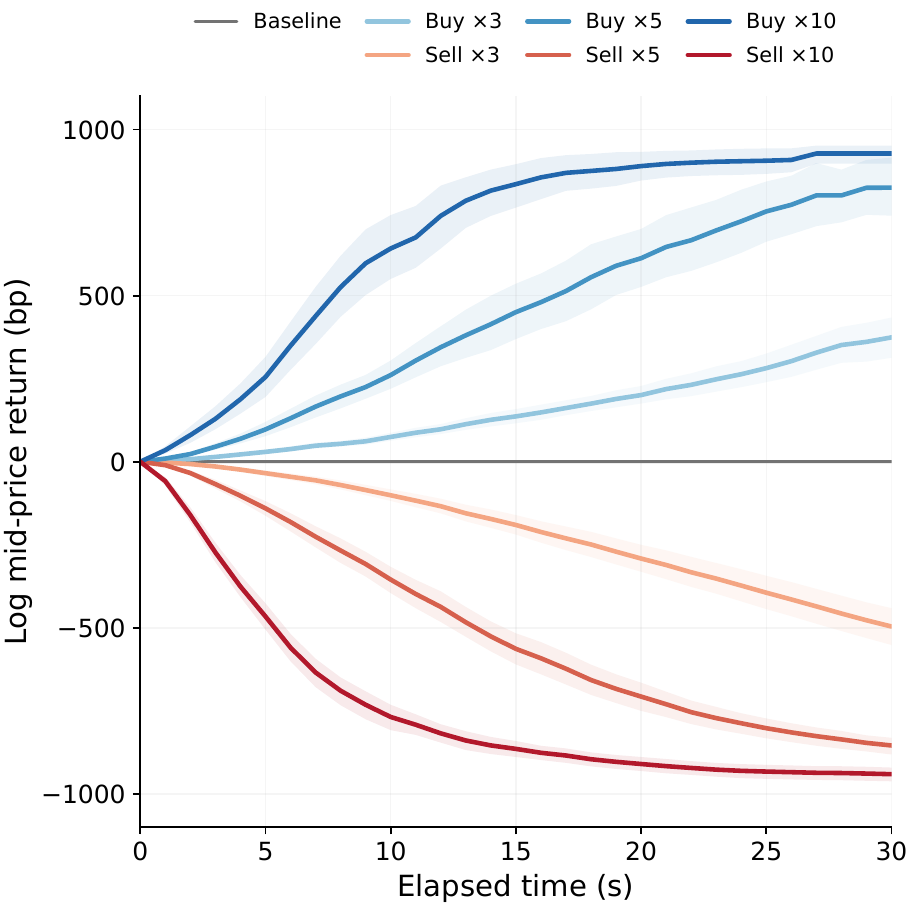}
    \caption{Counterfactual stress testing.}
    \label{fig:whatif_stress}
\end{subfigure}
\begin{subfigure}{0.56\linewidth}
    \centering
    \includegraphics[width=\linewidth]{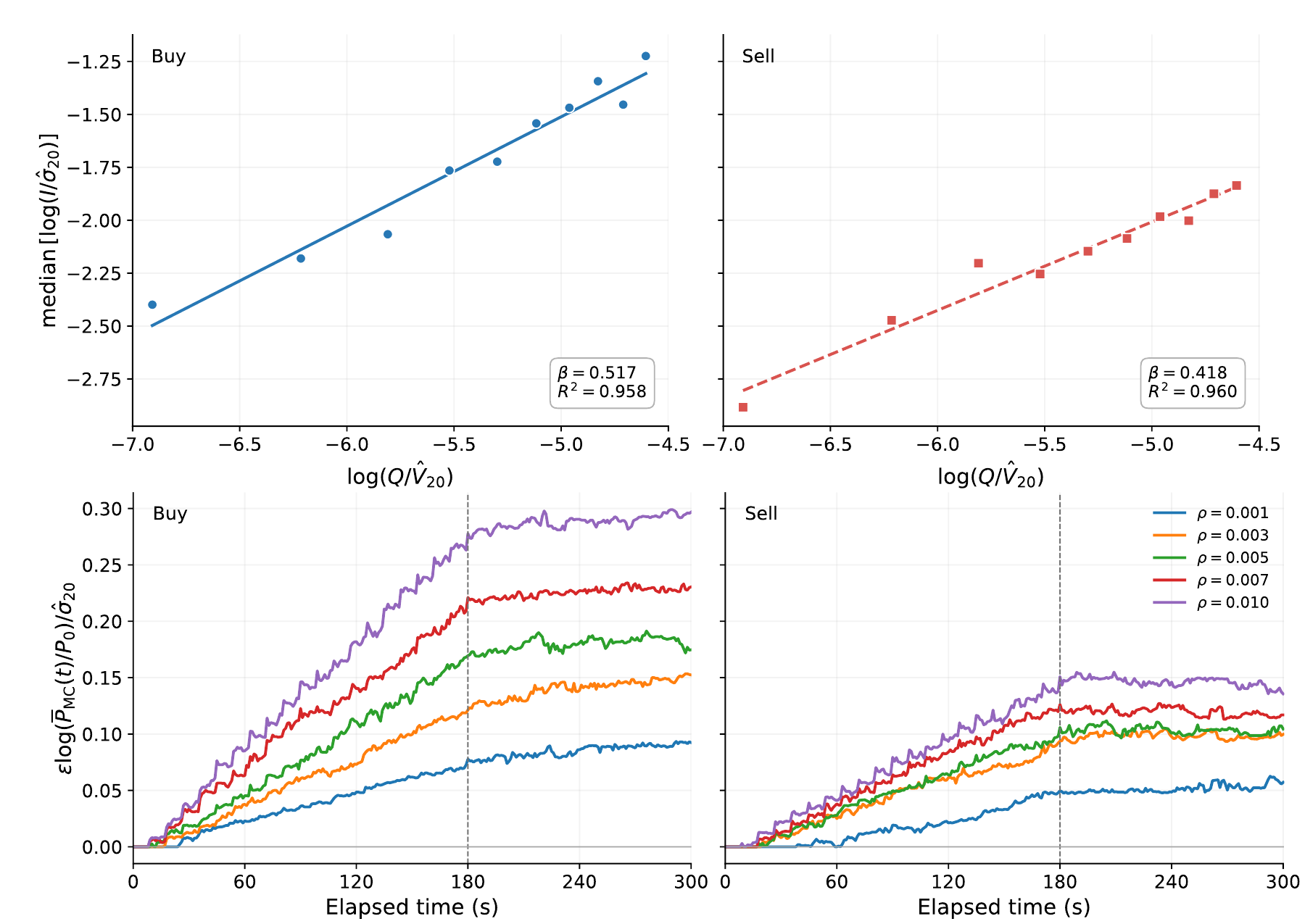}
    \caption{TWAP market impact.}
    \label{fig:whatif_sqrt}
\end{subfigure}
\caption{
\textbf{What-if analysis of the market microstructure model.}
Left: directional buy- and sell-side order-flow pressure produces
economically consistent mid-price responses. Right: TWAP
meta-order simulations exhibit concave square-root-like impact scaling with
respect to executed quantity.
}
\label{fig:whatif_analysis}
\end{figure*}


\paragraph{Volatility prediction.}
We forecast realized volatility by averaging the variances of 32 generated mid-price paths.
Figure~\ref{fig:volatility} reports the mean and median observed
realized volatility within quintiles of the variance forecasts.
Both statistics increase monotonically across quintiles at
1, 3, and 5 minutes, supporting the model's ability to rank
future volatility.

\subsection{What-if Analysis}

\paragraph{Intervention-based diagnostics.}
A market microstructure model should respond plausibly to controlled
interventions. Following the ``what-if'' evaluations in TradeFM and
MarS~\citep{kawawabeaudan2026tradefmgenerativefoundationmodel,
li2025marsfinancialmarketsimulation}, we consider two complementary tests:
a directional stress test that isolates the mechanical response
of the order book to one-sided order-flow pressure, and a TWAP
experiment that measures the price impact of meta-orders.

\paragraph{Directional stress testing.}
We generate continuations under an unmodified baseline and under
buy- and sell-side stress at several intensities. Buy stress amplifies
buy-side add events while leaving sell-side events unchanged, and sell
stress is defined symmetrically. Replicated events are matched and fed
back into the autoregressive context, allowing subsequent order flow
to respond to the intervention.

Figure~\ref{fig:whatif_stress} shows that buy stress raises the
mid-price relative to the baseline, whereas sell stress lowers it.
Stronger interventions generally produce faster and larger early
displacements. A possible explanation is that the abrupt increase in
same-side order-arrival intensity shifts the model toward a
high-activity, strongly imbalanced regime, encouraging continuations
resembling a market approaching a price limit. The precise intervention,
display, and aggregation procedures are given in the appendix.




\paragraph{Market impact and the square-root law.}
Empirically, the impact of a meta-order is commonly approximated by
\begin{equation}
I(Q) \approx Y \sigma_{i,d}\sqrt{Q/V_{i,d}},
\label{eq:square-root-law}
\end{equation}
where $Q$ is the meta-order size, $V_{i,d}$ and $\sigma_{i,d}$ are the
daily trading volume and realized volatility of security $i$ on day $d$,
and $Y$ is a scaling constant~\citep{moro2009market,toth2011anomalous,
bucci2019crossover}.

We examine this scaling using M3, the target \textbf{meta-order} sizes range from $0.10\%$ to $1.00\%$ of \textbf{estimated daily volume}. Both volume and volatility estimates are constructed as separate medians over the preceding 20 trading days.
Each simulation lasts 300 seconds: a TWAP meta-order is submitted through 20 child orders during the first 180 seconds, followed by 120 seconds without further intervention. Child orders are processed by the matching engine and encoded back into the autoregressive context, allowing subsequent background order sequences to condition on the intervention.

At the end of execution, we average the four Monte Carlo mid-price
predictions for each prompt and measure the direction-adjusted log return
relative to its initial mid-price. Figure~\ref{fig:whatif_sqrt} plots the
median of per-prompt log normalized impacts against log normalized executed
size. The estimated slopes are $0.517$ for buy meta-orders and $0.418$ for sell meta-orders, with $R^2$ values of $0.958$ and $0.960$, respectively. Both sides exhibit sublinear impact scaling.

\section{Conclusion}

We introduced \textbf{M3} (\underline{M}arket \underline{M}icrostructure \underline{M}odel), a market microstructure world model for order-level financial market simulation. \textbf{M3} represents heterogeneous order events and LOB states as discrete tokens, explicitly capturing the interaction between order events and market state, and then generated orders are recursively executed through the matching engine. This design allows the model to generate multiple future market trajectories.

Empirically, \textbf{M3} demonstrates strong and scalable order-flow modeling ability. Through empirical evaluation, we show that the generated trajectories reproduce key market properties, preserve short-horizon predictive information, and support counterfactual analysis such as liquidity stress testing and meta-order impact estimation. These results suggest that order-level generative world models can provide a useful foundation for simulation-based market analysis, risk evaluation, and strategy development. Future work may include incorporating richer cross-asset information and improving the context length of foundation models. We believe that microstructure models offer a promising direction for studying financial markets.

\bibliography{aaai2027}


\clearpage

\appendix

\section{Data and Preprocessing}
\label{appx:data_preprocessing}

\subsection*{Dataset Details}
The M3 training split spans January 2024 to November 2025 and contains approximately 31.9 billion order events. The evaluation uses December 2025 and January--February 2026. We preserve all limit and market orders in their original arrival order without separating them by order type. 

\begin{figure}[h]
    \centering
    \includegraphics[width=0.95\linewidth]{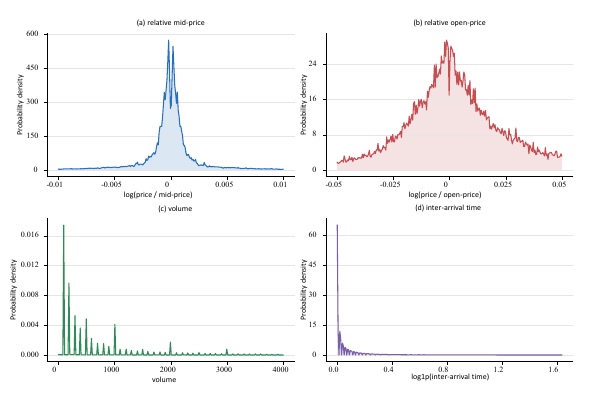}
    \caption{
    Empirical distributions of the continuous order-flow features. 
    Panels (a)--(d) show the distributions of relative mid-price, relative open-price, order volume, and inter-arrival time, respectively. 
    The features exhibit strong discreteness and heavy tails.
    }
    \label{fig:continuous_feature_distribution}
\end{figure}

\subsection*{Input Preprocessing}
Each order sequences window is denoted by $\mathbf{x}_{1:N}$, with
$\mathbf{x}_n\in\mathbb{R}^{5}$. The tokenizer uses the open-relative
price, log-volume $\log(1+v_n)$, inter-event time $\Delta\tau_n$,
action, and side. Only the three continuous features are clipped,
using fixed 0.5th and 99.5th percentile bounds. The same bounds are used during training and
encoding.

Following the data preprocessing pipeline, we restrict the sequence length to a maximum of 1024 tokens. 
Training data are partitioned into fixed, non-overlapping windows
of $N=1024$ order events within each stock and trading day.
The 1024-event length excludes the additional LOB-prefix and BOS
positions used by the autoregressive model.

\section{Tokenizer Details}
\label{appx:tokenizer}

\subsection{Tokenizer Architecture}

\paragraph{Network Design.} 
The tokenizer uses causal Transformer stacks for both encoding and
decoding, with a shared learned absolute positional embedding table.
Both stacks use Pre-Layer Normalization and GELU activations.
Their self-attention is
\begin{equation}
\operatorname{Attention}(\mathbf{Q},\mathbf{K},\mathbf{V})
=\operatorname{softmax}\!\left(
\frac{\mathbf{Q}\mathbf{K}^{\mathsf T}}{\sqrt{d_k}}
+\mathbf{M}\right)\mathbf{V},
\end{equation}
where $M_{n,m}=0$ for $m\leq n$ and $M_{n,m}=-\infty$ otherwise.
Thus, neither stack attends to future events. The decoder operates
on quantized latents and does not use encoder--decoder cross-attention.

\paragraph{Model Configuration.} 
The tokenizer has 11,266,951 (11M) trainable parameters across 3 encoder and 3 decoder layers, excluding the EMA-updated codebook of $4{,}194{,}304$ (4M) values. The multi-head self-attention module employs 6 independent heads with a hidden dimension of $d_{\text{model}} = 384$, while the feed-forward networks expand to $d_{\text{ff}} = 1536$. Designed to process sequences up to 1024 timesteps, the model linearly projects encoder outputs into a 128-dimensional bottleneck. Discretization is performed against a high-capacity learned codebook of $K=32,768$ latent embeddings, providing a robust vocabulary to capture complex order events.

\subsection{Tokenizer Training Objective}

\paragraph{Composite Objective Function.} 
To ensure both micro-structural fidelity and macroscopic statistical consistency in the reconstructed data, the model is optimized using a composite loss function, $L_{\text{tok}}$, defined as a weighted sum of four components:

\begin{equation}
\begin{aligned}
\mathcal{L}_{\mathrm{tok}}
={}&\mathcal{L}_{\mathrm{cont}}
+\mathcal{L}_{\mathrm{cat}}+\mathcal{L}_{\mathrm{vq}}\\
&+10^{-3}\mathcal{L}_{\mathrm{tick}}.
\end{aligned}
\end{equation}

\begin{itemize}
    \item \textbf{Continuous Reconstruction ($\mathcal{L}_{\text{cont}}$):} A weighted Mean Squared Error (MSE) applied to continuous features. Asymmetric dimensional weighting (e.g., $[100, 1, 1]$) is strictly applied to heavily prioritize the precise reconstruction of core price trajectories over auxiliary signals.
    \item \textbf{Categorical Reconstruction ($\mathcal{L}_{\text{cat}}$):} A standard Cross-Entropy (CE) loss to reconstruct discrete structural attributes, such as market actions and order sides.
    \item \textbf{Quantization Constraints ($\mathcal{L}_{\text{vq}}$):} The VQ term contains the commitment penalty with weight $\beta=0.25$ already included; no separate codebook-regression or code-usage regularizer is active.
    \item \textbf{Financial Heuristics Tick Loss ($\mathcal{L}_{\text{tick}}$):} A domain-specific Open Tick Loss ($\lambda_{\text{tick}} = 0.001$). This is a soft reconstruction penalty; exact tick alignment is obtained by rounding decoded prices before matching. We compute a Smooth L1 loss (Huber loss) between the continuous reconstructed tick values and the discrete target tick indices.
\end{itemize}

\paragraph{VQ-VAE Training Details.} 
We used a codebook of $K=32{,}768$ embeddings, each with dimension 128, and assigned each encoder output to its nearest code under Euclidean distance. The codebook was initialized using 10 iterations of $k$-means and updated by exponential moving averages with decay $0.99$. Codes with an EMA cluster size below 2 were replaced using current-batch encoder vectors. The VQ objective contained a commitment loss $\beta\lVert z_e-\operatorname{sg}(e_q)\rVert_2^2$ with $\beta=0.25$. Gradients through the hard assignment use the
rotation-trick estimator implemented in
\textit{vector-quantize-pytorch} v1.20.4.

We monitor code utilization as $U=|\{k:n_k>0\}|/K$ and
perplexity as $\exp[-\sum_k p_k\log p_k]$, where $p_k$ is the
empirical assignment frequency. At the selected checkpoint
(step 28,000), all 32,768 codes had been selected at least once,
giving 100\% cumulative utilization. Logged training microbatch
averages were 23,257 active codes (71.0\%) and perplexity 17,596;
the corresponding validation averages were 13,859 codes (42.3\%)
and perplexity 7,294. Cumulative coverage and microbatch activity
measure different properties: full cumulative coverage alone
does not rule out inactive codes or codebook collapse at a later step.

\subsection{Reconstruction Evaluation}\label{appx:recons}
Table~\ref{tab:appendix_reconstruction_comparison} evaluates order-event reconstruction under a common vocabulary budget of $K=32768$. For all tokenizers, we keep the original order sequence fixed: each original event is encoded and decoded, and the decoded event is then submitted to the matching engine to construct a reconstructed order-book trajectory.

The two bin results differ in the price anchor available during decoding. Bin-O is a privileged oracle diagnostic: each event is decoded relative to the contemporaneous mid-price from the original order-book trajectory. Although this anchor is available in an offline reconstruction experiment, it is unavailable during autoregressive generation, because the ground-truth order-book trajectory is then unknown. Bin-S instead uses the mid-price of the reconstructed book obtained by replaying all previously decoded events through the matching engine. It therefore reproduces the anchoring mechanism that must be used at generation time. Small price or volume reconstruction errors can alter the reconstructed book, shift the anchor used for subsequent events, and accumulate along the trajectory. The difference between Bin-O and Bin-S consequently measures the sensitivity of the relative-mid tokenizer to this recursive anchor feedback; Bin-O should not be interpreted as a generation-feasible baseline.

Our VQ tokenizer avoids this feedback mechanism by representing prices relative to the daily open price, which is fixed and available throughout both reconstruction and generation. It additionally learns a compact discrete codebook over the continuous order features. Among the generation-feasible decoding schemes, VQ outperforms Bin-S on every reported price metric, reducing the price MAE from \num{4.3e-1} to \num{2.51e-2}, increasing both exact-tick and within-one-tick accuracy, and reducing the P90 and P99 errors from 26 and 764 ticks to 6 and 28 ticks, respectively. It also consistently improves the reconstruction of the other continuous features. These results motivate VQ as the discretization interface for autoregressive order-flow modeling.

\begin{table}[t]
\centering
\vspace{0.35em}
\small
\setlength{\tabcolsep}{8pt}
\renewcommand{\arraystretch}{1.05}
\begin{subtable}[t]{\columnwidth}
\centering
\setlength{\tabcolsep}{3pt}
\begin{tabular}{@{}lccc@{}}
\toprule
& \multicolumn{1}{c}{\textbf{Oracle diagnostic}}
& \multicolumn{2}{c}{\textbf{Generation-feasible}} \\
\cmidrule(lr){2-2}
\cmidrule(lr){3-4}
\textbf{Metric}
& \textbf{Bin-O$^\dagger$}
& \textbf{Bin-S}
& \textbf{VQ} \\
\midrule
Price MAE
& \num{5.86e-2}
& \num{4.3e-1}
& \bfseries \num{2.51e-2} \\

\addlinespace[0.12em]
Exact tick rate
& 0.765
& 0.307
& \bfseries 0.377 \\

Within 1 tick
& 0.846
& 0.548
& \bfseries 0.662 \\

\addlinespace[0.12em]
Err. P90 (ticks)
& 4
& 26
& \bfseries 6 \\

Err. P99 (ticks)
& 123
& 764
& \bfseries 28 \\
\bottomrule
\end{tabular}
\parbox{0.94\linewidth}{\footnotesize
$^\dagger$Bin-O 
is unavailable
during generation and must not be interpreted as a
deployable baseline. Bold denotes the better result among generation-feasible methods.
}
\caption{
Price reconstruction under oracle and generation-time information
settings (↓ MAE/P90/P99, ↑ hit rates).
Bin-O uses the ground-truth contemporaneous mid-price and is included
only as a non-causal diagnostic; the deployable comparison is Bin-S
versus VQ.
}
\label{tab:appendix_price_reconstruction_anchor}
\end{subtable}

\vspace{0.80em}

\begin{subtable}[t]{0.85\columnwidth}
\centering

\setlength{\tabcolsep}{4pt}
\begin{tabular}{@{}lcc@{}}
\toprule
\textbf{Metric}
& \textbf{Bin tokenizer}
& \textbf{VQ tokenizer} \\
\midrule
Relative price MAE   & \num{1.119e-3} & \bfseries \textbf{\num{5.250e-4}}\\
Log-volume MAE       & \num{6.910e-2}  & \bfseries \num{3.530e-2} \\
Volume MAE           & \num{5.064e2} & \bfseries \num{1.171e2} \\
Delta-time MAE       & \num{2.380e-2}  & \bfseries \num{1.260e-2} \\
Event-time MAE       & \num{4.443e0}  & \bfseries \num{1.553e0} \\
Final-time abs. err. & \num{8.616e0}  & \bfseries \num{2.770e0} \\
\bottomrule
\end{tabular}
\caption{Reconstruction accuracy of other continuous features (↓ better).}
\label{tab:appendix_feature_reconstruction}
\end{subtable}

\vspace{0.45em}

\begin{minipage}{0.86\columnwidth}
\footnotesize
\emph{Notes.}
Bold indicates the best value in each row. Lower is better for error metrics; higher is better for rate metrics.
``Bin (oracle)'' uses the true current mid-price anchor, while ``Bin (sim.)'' uses a simulated mid-price anchor.
The bin tokenizer decodes each token by the median value of its bin.
\end{minipage}
\caption{
Reconstruction accuracy comparison under a fixed vocabulary size of $K=32768$.
For the bin-based tokenizer, the vocabulary is factorized into
$32$ price bins, $16$ volume bins, $16$ time bins, $2$ action types, and $2$ sides.
}
\label{tab:appendix_reconstruction_comparison}
\end{table}

\section{Autoregressive Model Details}
\label{appx:ar_model}

\subsection{Autoregressive Backbone}


We adopt a decoder-only LLaMA2-style Transformer as the
autoregressive backbone for modeling discrete order
tokens~\citep{vaswani2017attention,touvron2023llama}.
An order sequence window is represented by
$C=(c_1,\ldots,c_N)$, where $N$ is the number of
order tokens and $c_n\in\{0,\ldots,K-1\}$.
The order-code vocabulary contains $K=32{,}768$ entries.
With three additional special tokens for padding,
beginning of sequence (BOS), and end of sequence (EOS),
the autoregressive vocabulary has size
$K_{\mathrm{AR}}=K+3$.

Let $\mathbf{r}^{\mathrm{lob}}_{0,1:P}$ denote the continuous
LOB prefix embeddings for the initial window $k=0$, and let
$c_0=c_{\mathrm{BOS}}$ denote the start token.
The model learns the conditional factorization
\begin{equation}
p_{\theta_{\mathrm{AR}}}
\left(C\mid\mathbf{r}^{\mathrm{lob}}_{0,1:P}\right)
=
\prod_{n=1}^{N}
p_{\theta_{\mathrm{AR}}}
\left(
c_n\mid
\mathbf{r}^{\mathrm{lob}}_{0,1:P},c_0,c_{<n}
\right).
\end{equation}
Thus, each order token is predicted from the initial market
state and preceding order tokens.

The order-token embeddings are concatenated with the LOB
prefix and BOS embedding, then processed by a stack of
$L_{\mathrm{dec}}$ decoder layers.
Each layer uses pre-normalized causal multi-head
self-attention with rotary positional embeddings, followed
by a gated feed-forward network with SiLU
activation~\citep{su2023roformerenhancedtransformerrotary,
zhang2019rmsnorm,shazeer2020glu}.
We use standard multi-head attention.
After the final decoder layer, RMSNorm and a linear
language-modeling head produce vocabulary logits:
\begin{equation}
\begin{aligned}
\mathbf{O}
&=
\mathrm{LMHead}
\left(
\mathrm{RMSNorm}
\left(\mathbf{H}^{(L_{\mathrm{dec}})}\right)
\right),\\
\mathbf{O}
&\in
\mathbb{R}^{B\times(P+1+N)\times K_{\mathrm{AR}}},
\end{aligned}
\end{equation}
where $B$ is the batch size.
Here, $N$ counts only order tokens; the LOB prefix
and BOS token occupy additional input positions.

Training minimizes the mean next-token negative log-likelihood:
\begin{equation}
\mathcal{L}_{\mathrm{AR}}
=
-\frac{1}{N}\sum_{n=1}^{N}
\log p_{\theta_{\mathrm{AR}}}
\left(
c_n\mid
\mathbf{r}^{\mathrm{lob}}_{0,1:P},c_0,c_{<n}
\right).
\end{equation}
Only order-token targets contribute to the loss;
LOB-prefix and BOS positions are excluded.

\begin{figure}[ht]
  \centering
  \includegraphics[width=0.90\linewidth]
  {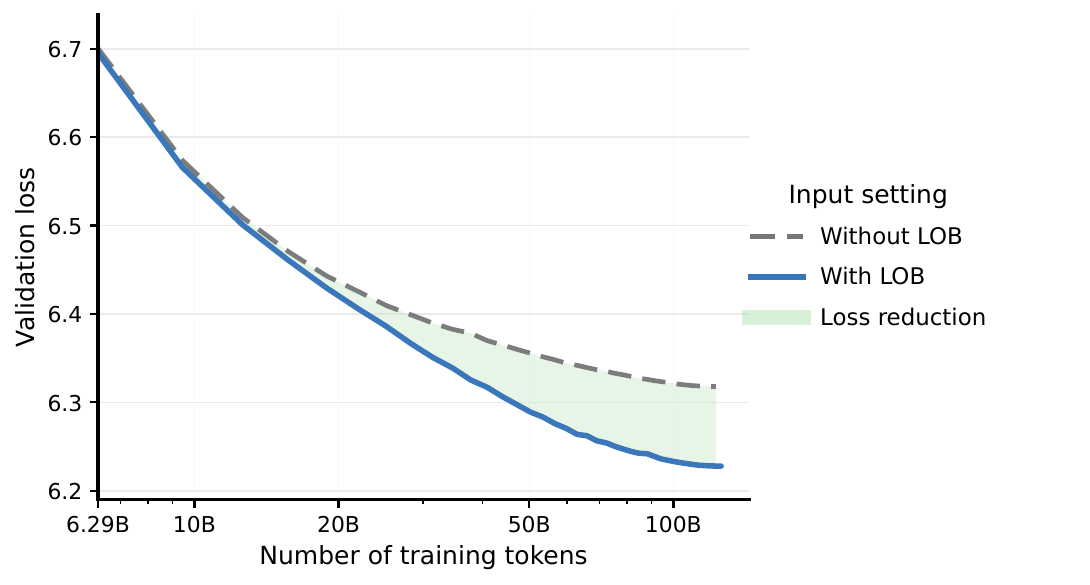}
  \caption{LOB loss comparison.}
  \label{fig:lob_loss_comparison}
\end{figure}

\subsection{LOB Prefix Encoder}

For window $k$, let $L_k$ denote the LOB snapshot immediately
before its first event. The initial window corresponds to $k=0$.
This snapshot describes the liquidity state from which
the subsequent order sequence evolves.
The LOB prefix encoder converts this state into continuous
embeddings that condition autoregressive order prediction.

For each sample, the snapshot is represented as
$L_k\in\mathbb{R}^{2\times d_L\times2}$, with $d_L=100$.
The dimensions index book side, price level, and feature
(absolute price or volume), respectively.
Let $p^{(k)}_{b,\ell}$ and $v^{(k)}_{b,\ell}$ denote the price and
volume at side $b\in\{\mathrm{ask},\mathrm{bid}\}$ and
level $\ell\in\{1,\ldots,d_L\}$.
Using the reference mid-price $m_k$
associated with the snapshot, we construct
\begin{equation}
\mathbf{u}^{(k)}_{b,\ell}
=
\begin{bmatrix}
\operatorname{clip}\!\left(
\dfrac{p^{(k)}_{b,\ell}-m_k}{m_k},-0.20,0.20\right)\\
\log\!\left(1+v^{(k)}_{b,\ell}\right)
\end{bmatrix}
\in\mathbb{R}^{2},
\end{equation}
where $0.20$ is the price clipping threshold.


Each level feature is projected into the autoregressive
hidden dimension $d_{\mathrm{AR}}$.
We add learned side and level embeddings, together with
a market-state vector $\mathbf{s}_k$ containing normalized spread,
intraday time divided by 14,400 seconds, and return from the open.
Spread and return are clipped to $[-0.2,0.2]$, and normalized time
to $[0,1]$. A two-layer MLP with GELU projects this vector:
\begin{equation}
g_{\mathrm{state}}(\mathbf{s}_k)
=W_2\operatorname{GELU}(W_1\mathbf{s}_k+\mathbf{b}_1)
+\mathbf{b}_2.
\end{equation}
The input embedding for each book position is
\begin{equation}
\begin{aligned}
\mathbf{h}^{(0)}_{k,b,\ell}
={}&
W_{\mathrm{lob}}\mathbf{u}^{(k)}_{b,\ell}+\mathbf{b}_{\mathrm{lob}}
+\mathbf{e}^{\mathrm{side}}_b\\
&+\mathbf{e}^{\mathrm{level}}_\ell
+g_{\mathrm{state}}(\mathbf{s}_k).
\end{aligned}
\end{equation}
This construction distinguishes all
$2d_L=200$ book positions while sharing parameters
across sides and levels.

The resulting embeddings are stacked into
$\mathbf{H}^{(0)}_{\mathrm{LOB},k}
\in\mathbb{R}^{2d_L\times d_{\mathrm{AR}}}$
and processed by a lightweight Transformer encoder
with two layers and four attention
heads~\citep{vaswani2017attention}:
\begin{equation}
\begin{aligned}
\mathbf{H}_{\mathrm{LOB},k}
&=
\mathrm{TransformerEncoder}
\left(\mathbf{H}^{(0)}_{\mathrm{LOB},k}\right),\\
\mathbf{H}_{\mathrm{LOB},k}
&\in\mathbb{R}^{2d_L\times d_{\mathrm{AR}}}.
\end{aligned}
\end{equation}

Finally, cross-attention pooling compresses the
$2d_L=200$ encoded book positions into $P=32$ prefix embeddings.
A set of learned queries
$\mathbf{Q}_{\mathrm{lob}}
\in\mathbb{R}^{P\times d_{\mathrm{AR}}}$
attends to the encoded book states as keys and values:
\begin{equation}
\begin{aligned}
\mathbf{r}^{\mathrm{lob}}_{k,1:P}
&=
G_{\psi}(L_k,\mathbf{s}_k)\\
&=
\mathrm{LayerNorm}
\Bigl(
\mathrm{MHA}
\bigl(
\mathbf{Q}_{\mathrm{lob}},
\mathbf{H}_{\mathrm{LOB},k},
\mathbf{H}_{\mathrm{LOB},k}
\bigr)
\Bigr),\\
\mathbf{r}^{\mathrm{lob}}_{k,1:P}
&\in\mathbb{R}^{P\times d_{\mathrm{AR}}}.
\end{aligned}
\end{equation}
These continuous prefix embeddings summarize the current
liquidity state and are prepended to the order-token
sequence.





\subsection{Matching Engine}
\label{app:matching_engine}

We use a deterministic matching engine to translate generated order
events into mid-price trajectories. The engine maintains bid and ask
books, internal order queues, and a simulation clock. 
Matching retains all supplied and subsequently created price levels.
The following describes the continuous-trading configuration used
in our experiments.

\paragraph{Book initialization.}
The engine is initialized from the observed LOB snapshot preceding
the historical prompt. Each occupied side--price pair is represented
by one synthetic resting order carrying its observed aggregate volume.
This preserves the snapshot depths without recovering the original
order identities or within-level queue composition.
Observed prompt events are then replayed in sequence to establish
the state at the forecast origin. Thereafter, the book evolves through
generated events, with its state preserved across generation windows.

\paragraph{Event legalization.}
For decoded event \(n\), let \(\widehat{\rho}_n\) denote the relative
order price, \(\widehat{\ell}_n\) the log-volume output corresponding
to \(\log(1+v)\), and \(\widehat{\Delta \tau}_n\) the inter-event interval in seconds.
Using the fixed opening mid-price \(m_{\mathrm{open}}\) as the price
anchor, we construct
\begin{equation}
\begin{aligned}
p_n
&=
\delta\,
\operatorname{round}\!\left(
\frac{m_{\mathrm{open}}(1+\widehat{\rho}_n)}{\delta}
\right),\\
v_n
&=
q\,\operatorname{round}\!\left(
\frac{\exp(\widehat{\ell}_n)-1}{q}
\right),\\
\tau_n
&=
\tau_{n-1}+\widehat{\Delta\tau}_n.
\end{aligned}
\label{eq:order_legalization}
\end{equation}
Here, \(\delta=0.01\) CNY is the price tick,
\(q=1\) share is the quantity grid.
Rounding uses the nearest integer, with ties to even.
These fixed numerical bounds are applied before matching.

The decoded action is rounded to an integer: zero denotes cancellation
and a nonzero value denotes addition. The decoded side is rounded and
clamped to \(\{0,1\}\), with one denoting buy and zero denoting sell.
A decoded block containing non-finite values terminates the
corresponding rollout before that block is replayed.

\paragraph{Price--time matching.}
Matching follows price–time priority. Each order gets an internal engine ID. A buy consumes asks from low to high, and a sell consumes bids from high to low, within the limit price; at each price, resting orders are filled FIFO. Each trade is for the smaller of the incoming remainder and the queue-head remainder, at the resting price. Any leftover joins the queue at its original limit. Cancels specify only side, price, and quantity, so the engine removes quantity FIFO at that side and price, capped by available depth; shortfalls are logged and do not affect other levels. Internal IDs are for accounting only.


\paragraph{Clock and mid-price updates.}
The nonnegative inter-event intervals advance the simulation clock.
Events sharing a timestamp retain their sequence order, and
timestamp-level price sampling uses the final book state.

\subsection{Model Hyperparameters}\label{appx:model_parameters}

Our final hyperparameters are shown in Table~\ref{tab:model_hyperparameters}.

\begin{table}[t]
\centering
\small
\setlength{\tabcolsep}{3pt}
\begin{tabular}{lccccc}
\toprule
\textbf{Model} & \textbf{Size} & \textbf{Hidden} & \textbf{FFN} & \textbf{Layers} & \textbf{Heads} \\
\midrule
tiny   & 10M   & 128  & 512  & 3  & 4  \\
small  & 25M   & 256  & 1024 & 6  & 4  \\
base   & 75M   & 512  & 2048 & 8  & 8  \\
large  & 366M  & 1024 & 4096 & 16 & 16 \\
xlarge & 1.27B & 2048 & 5504 & 20 & 16 \\
\bottomrule
\end{tabular}
\caption{Transformer hyperparameters for each model scale. All configurations use a vocabulary size of 32,768 and untied input/output embeddings.}
\label{tab:model_hyperparameters}
\end{table}

\section{Evaluation Details}
\label{appx:evaluation_details}

\subsection{Scaling Law}
\paragraph{Held-out validation and bootstrap uncertainty.}
We assessed the predictive validity of scaling law using the complete
$5\times4$ grid of 20 observed losses, comprising five model sizes
($9.66$M, $24.99$M, $74.75$M, $366.06$M, and $1.27$B parameters) and four
training-token budgets ($1$B, $3$B, $10$B, and $31.9$B tokens). For each
leave-one-model-size-out fold, we jointly removed all four observations
corresponding to one model size, refitted the quation to the
remaining 16 observations using the same Huber objective on log-loss residuals
($\delta=10^{-3}$), and predicted the four held-out losses. We define the
signed relative error as
$100(\widehat{L}-L)/L$ and report its absolute value averaged over the four
token budgets as the fold MAPE. The MAPEs for held-out sizes $9.66$M,
$24.99$M, $74.75$M, $366.06$M, and $1.27$B were $1.14\%$, $0.40\%$,
$0.56\%$, $0.70\%$, and $0.53\%$, respectively, giving an overall mean
absolute relative error of $0.66\%$ across the 20 held-out predictions. 

To quantify finite-sample uncertainty, we additionally performed a
fixed-design residual bootstrap with 2,000 replicates. Specifically, after
fitting the equation to all 20 observations, we computed and centered
the log-loss residuals
$\epsilon_i=\log L_i-\log\widehat{L}_i$. For each replicate, residuals were
sampled with replacement and pseudo-observations were generated as
$L_i^{*}=\widehat{L}_i\exp(\epsilon_i^{*})$ at the original $(M_i,D_i)$ design
points. The same Huber objective was then refitted to each pseudo-dataset.
Percentile intervals were obtained directly from the resulting parameter
distributions. The 90\% bootstrap confidence intervals were
$\alpha=0.379\,[0.281,\,0.485]$,
$\beta=0.312\,[0.211,\,0.423]$, and
$E=5.370\,[5.065,\,5.529]$; the corresponding 95\% intervals were
$\alpha=0.379\,[0.259,\,0.510]$,
$\beta=0.312\,[0.193,\,0.456]$, and
$E=5.370\,[4.972,\,5.562]$.
These intervals quantify residual uncertainty conditional on the observed
$(M,D)$ grid and the assumed scaling-law form; they do not capture run-to-run
training stochasticity because only one loss observation is available at each
configuration.

\subsection{Stylized Facts}
\label{appx:stylized_facts}

We compare empirical mid-price paths with those obtained by replaying
generated orders through the matching engine. 
M3 results in this section use the Large model. Realized zero returns
are retained, with no return-magnitude or book-depth exclusions.
For mid-price $m_t$, the horizon-$\Delta$ log return is
$r_{t,\Delta}=\log(m_{t+\Delta}/m_t)$.

\paragraph{Distributional fidelity.}
Panel A of Table~\ref{tab:tradefm_fidelity_with_kurtosis} compares
event-anchored rolling returns at
$\Delta\in\{10,20,60,120\}\,\mathrm{s}$.
Distances are computed separately for each stock, trading date,
and Monte Carlo index, pooling the daily forecast origins within
each comparison. For empirical and generated return CDFs $F$ and $G$,
we use the Kolmogorov--Smirnov statistic
$D_{\mathrm{KS}}(F,G)=\sup_x|F(x)-G(x)|$ and the Wasserstein distance
$W_1(F,G)=\int_0^1|F^{-1}(u)-G^{-1}(u)|\,\mathrm{d}u$.
The K--S statistic is dimensionless; $10^4W_1$ is reported in basis
points. Lower values indicate closer distributional agreement.

\paragraph{Return distributions and kurtosis.}
Figure~\ref{fig:stylized_facts}(a) and Panel B of
Table~\ref{tab:tradefm_fidelity_with_kurtosis} use origin-to-horizon
returns at $\Delta\in\{1,3,5\}\,\mathrm{min}$.
Pearson kurtosis is
$\operatorname{Kurt}(r)=\langle(r-\bar r)^4\rangle/
\langle(r-\bar r)^2\rangle^2$, where $\bar r=\langle r\rangle$
and brackets denote the sample average. Its Gaussian reference
value is three. Winsorized kurtosis clips each return pool at its
0.5th and 99.5th percentiles before recomputing the mean and central
moments. The table reports both raw and winsorized kurtosis;
figure annotations use winsorized values, while the plotted
densities use the original returns.

\paragraph{Return autocorrelation.}
Figure~\ref{fig:stylized_facts}(b) uses non-overlapping 20-second
log returns along each five-minute path, starting from the observed
origin mid-price. At lags $\ell\in\{0,20,\ldots,120\}\,\mathrm{s}$,
we compute $C_r(\ell)=\operatorname{Corr}(r_t,r_{t+\ell})$ and
$C_{|r|}(\ell)=\operatorname{Corr}(|r_t|,|r_{t+\ell}|)$, with
$r_t=r_{t,20\,\mathrm{s}}$.
Pearson correlations pool finite pairs from within the same path;
pairs never cross path boundaries. No winsorization is applied.
Lag zero is included with correlation one. Pooled absolute-return
correlations can also reflect volatility differences across origins.

\paragraph{Compared market simulators.}
ZI and Hawkes retain parameters calibrated on the training events
from November 24--28, 2025. They receive the same observed order
prompts and use the same matching engine as M3.
Write an event as $e_n=(a_n,d_n,\zeta_n,v_n,\Delta\tau_n,)$,
where $\Delta\tau_n$ is the inter-event time, $a_n$ the add/cancel
action, $d_n$ the side, and $v_n$ the volume. The signed price
displacement is $\zeta_n=10^4(p_n/m_{\tau_n^-}-1)$, measured
relative to the prevailing pre-event mid-price.

\paragraph{Zero-Intelligence baseline.}
ZI draws successive events independently, with calibrated marginal
distributions for their attributes:
\begin{equation}
p_{\mathrm{ZI}}(e_n)
=p(\Delta\tau_n)\,p(a_n)\,p(d_n)\,p(\zeta_n)\,p(v_n).
\label{eq:zi_factorization}
\end{equation}
Action and side follow categorical distributions fitted to their
empirical frequencies. Inter-event time and volume follow exponential
distributions with calibrated rates. Price displacement follows a
five-component Gaussian mixture,
$p(\zeta)=\sum_{g=1}^{5}\omega_g
\mathcal N(\zeta;\mu_g,\sigma_g^2)$.

\paragraph{Compound Hawkes baseline.}
Hawkes distinguishes four event types: buy-cancel, buy-add,
sell-cancel, and sell-add. Let $\kappa_n$ denote the type of event
$n$ and $\mathcal H_t$ the history before time $t$.
Its conditional intensities use mixtures of exponential kernels:
\begin{equation}
\begin{aligned}
\lambda_\kappa(t\mid\mathcal H_t)
&=\mu_\kappa+\sum_{n:\tau_n<t}
h_{\kappa,\kappa_n}(t-\tau_n),\\
h_{\kappa\nu}(u)
&=\sum_{q=1}^{4}\alpha_{\kappa\nu q}\beta_q
e^{-\beta_q u},\qquad u>0.
\end{aligned}
\label{eq:compound_hawkes_intensity}
\end{equation}
Here $\mu_\kappa$ is the background intensity and
$\alpha_{\kappa\nu q}$ the integrated excitation from type $\nu$
to type $\kappa$ at scale $q$. Decay rates are
$\beta_q=\log 2/\tau_q^{\mathrm{half}}$, with half-lives
$\tau_q^{\mathrm{half}}\in\{0.05,0.5,5,60\}\,\mathrm{s}$.
The observed order prompt initializes the excitation state;
subsequent events are sampled by thinning.
Price-displacement marks use type-specific five-component Gaussian
mixtures, and volume marks use type-specific exponential distributions.

\subsection{Mid-price Prediction}
\label{app:midprice_prediction}


\paragraph{Prompt and rollout.}
The initial prompt consists of an observed LOB snapshot followed by up
to 512 historical order-event tokens ending at the forecast origin.
The snapshot precedes this historical order block. The matching engine
is initialized from the corresponding full-depth book and replays the
historical orders before the generated continuation. Neural LOB
encoder uses 100 price levels per side, while matching retains
all available price levels.

Each autoregressive segment accommodates 1,024 order tokens, including
the historical tokens in the first segment. Generated orders are
decoded, legalized and processed sequentially by the matching engine. At a segment boundary, the updated LOB and state condition a fresh segment with a start token; previous order tokens are discarded. The matching state and simulation clock persist across segments. We generate $S=32$ continuations per prompt, stopping after completing the event timestamp cluster covering the five-minute forecast boundary.

\paragraph{Horizon prices and returns.}
Let $i$ index stocks and $t$ denote a forecast origin. The origin mid-price is $m_{i,t}$. At horizon $h$, each
generated price $\widehat m^{(s)}_{i,t}(h)$ and the observed price
$m_{i,t}(h)$ denote mid-prices selected at the event timestamp nearest to $t+h$.
Observed targets are taken from the empirical order-book path. Undefined selected quotes remain undefined, with no additional time-mismatch threshold.
The ensemble price forecast is
$\widehat m_{i,t}(h)=S^{-1}\sum_{s=1}^{S}
\widehat m^{(s)}_{i,t}(h)$.
We use this price mean to compute the predicted simple return:
 \begin{equation}
 \begin{aligned}
\widehat R_{i,t,h}
&=\frac{\widehat m_{i,t}(h)}{m_{i,t}}-1,\\
R_{i,t,h}
&=\frac{m_{i,t}(h)}{m_{i,t}}-1.
\end{aligned}
\label{eq:return_forecasts}
 \end{equation}



\paragraph{Directional accuracy evaluation.}
Figure~\ref{fig:midprice_prediction} evaluates M3 Large 
on the ability of identifying the directionality of market advances and declines. The accuracy curve uses horizons of 30 seconds and
1, 3, and 5 minutes. 
At each horizon, valid prompts in the directional panel are ranked
globally by predicted returns and divided into lower, middle, and
upper thirds, labeled Down, Flat, and Up. Realized terciles are
constructed independently from observed returns.
A fixed prompt-identifier order resolves ties. Flat denotes the
middle tercile, rather than a zero-return class.

Let $\mathcal D_h$ and $\mathcal U_h$ denote the lower and upper
predicted terciles. After forming these groups, define
\begin{equation}
\mathcal E_h=
\left\{(i,t)\in\mathcal D_h\cup\mathcal U_h:
R_{i,t,h}\ne0\right\}.
\end{equation}

Assign $\widehat d_{i,t,h}=-1$ in $\mathcal D_h$ and $+1$ in
$\mathcal U_h$. Tail directional accuracy is
\begin{equation}
\mathrm{Acc}_{\mathrm{tail}}(h)
=\frac{1}{|\mathcal E_h|}
\sum_{(i,t)\in\mathcal E_h}
\mathbf{1}\!\left\{\widehat d_{i,t,h}R_{i,t,h}>0\right\}.
\label{eq:tail_accuracy}
\end{equation}

An independent, equally likely positive or negative sign has expected
accuracy $1/2$ on this nonzero-return subset, irrespective of its
realized sign balance. The diagnostic measures directional ranking
rather than the profitability of an executable trading strategy.



\paragraph{Confusion matrices.}
Both matrices in Figure~\ref{fig:midprice_prediction}(b,c) use the
one-minute horizon. The sign matrix uses $\mathcal E_h$ and compares
predicted tail directions with realized return signs. The tercile
matrix compares predicted and realized terciles on all valid prompts,
including realized zero returns. Counts are normalized within each
predicted-label row. Under independent predicted and realized tercile
assignments, the expected row fractions are approximately one third.



\subsection{Return Prediction}
\label{app:return_prediction}

Let $\mathcal P_h\subseteq\mathcal I\times\mathcal T$ contain the valid
stock--origin pairs at horizon $h$. 
Let $N_h=|\mathcal P_h|$. Define the grand means
\begin{equation}
\begin{aligned}
\overline{\widehat R}_h
&=N_h^{-1}\sum_{(i,t)\in\mathcal P_h}\widehat R_{i,t,h},\\
\overline R_h
&=N_h^{-1}\sum_{(i,t)\in\mathcal P_h}R_{i,t,h}.
\end{aligned}
\end{equation}
Writing $a_{i,t,h}=\widehat R_{i,t,h}-\overline{\widehat R}_h$ and
$b_{i,t,h}=R_{i,t,h}-\overline R_h$, the panel Pearson IC is
\begin{equation}
\mathrm{IC}_{\mathrm{panel}}(h)
=\frac{\displaystyle\sum_{(i,t)\in\mathcal P_h}a_{i,t,h}b_{i,t,h}}
{\displaystyle
\sqrt{\sum_{(i,t)\in\mathcal P_h}a_{i,t,h}^{2}}\,
\sqrt{\sum_{(i,t)\in\mathcal P_h}b_{i,t,h}^{2}}}.
\label{eq:panel_ic_expanded}
\end{equation}
RankIC applies the same formula after replacing both return series
by their global ranks within $\mathcal P_h$, with average ranks for
ties.

\paragraph{Microstructure Signal Baselines.}
At forecast origin $t$, let $b_t,a_t$ be the best bid and ask,
$q_t^b,q_t^a$ their displayed sizes, and $m_t=(a_t+b_t)/2$.
We compute queue imbalance~\citep{gould2016queue} and relative
weighted-mid displacement as
\begin{equation}
I_t=\frac{q_t^b-q_t^a}{q_t^b+q_t^a},
\qquad
W_t=I_t\frac{a_t-b_t}{2m_t}.
\end{equation}

For depth-normalized level-1 order flow imbalance
(OFI)~\citep{cont2014impact}, we compare successive book states after
applying all updates at each timestamp in $(t-150\,\mathrm{s},t]$.
On each side, an improved or newly appearing best quote contributes
its new size, an unchanged quote contributes the change in size, and
a worsened or disappearing quote contributes the negative of its
previous size. We sum bid contributions minus ask contributions
and divide by the time-weighted average of $q^b+q^a$ over the window.
All three scores are used directly at each forecast horizon without fitting.

\paragraph{Foundation Model Baselines.}
The baselines receive 10 completed 15-second midpoint bars ending at
the forecast origin, covering 150 seconds. Bars are constructed solely
from observed history; future targets are not exposed to inference.
Chronos-2, Toto, and Kronos receive midpoint OHLC features;
TimesFM receives midpoint closes. 
The 1-, 2-, and 3-minute forecasts are read at steps 4, 8, and 12 of
the native 15-second forecast grid, without interpolation.

All point signals use MC32 price means followed by
Equation~\ref{eq:return_forecasts}. Toto use native
stochastic samples, and Kronos uses its native MC32 ensemble output.
For Chronos-2 and TimesFM-2.5, the MC32 estimator approximates marginal forecast distributions from quantiles at levels $0.1,\ldots,0.9$. Quantile crossings are repaired by a cumulative maximum, followed by linear inverse-CDF interpolation with endpoint-clamped tails.
These draws are marginal approximations and are not interpreted as
native joint sample paths.

\paragraph{Fine-tuning.}
The four general-purpose models are evaluated with both pretrained
and November-only fine-tuned weights. Training uses November 3--21,
2025, and validation uses November 24--28. December observations are
excluded from local fine-tuning and checkpoint selection. The same
selected weights are used on both parts of the December panel.





\subsection{Volatility Prediction}
\label{app:volatility_calibration}

We reuse the expanded December panel of 256 stocks, eight dates, and
four daily origins (10:00, 10:30, 14:00, and 14:30), giving 8,192 prompts.
For each prompt, we use all $S=32$ saved M3 Large trajectories.
Generated prices were obtained with the matching engine.

Each path is sampled on a one-second previous-tick grid,
$u_j=j\delta$, with $\delta=1\,\mathrm{s}$.
At repeated event timestamps, we use the final book state.
For stock $i$, forecast origin $t$, path $s$, and horizon
$h\in\{60,180,300\}\,\mathrm{s}$, realized variance is
\begin{equation}
\mathrm{RV}^{(s)}_{i,t,h}
=\sum_{j=1}^{h/\delta}
\left[\log
\frac{\widetilde m^{(s)}_{i,t}(u_j)}
{\widetilde m^{(s)}_{i,t}(u_{j-1})}\right]^2.
\end{equation}
The forecast and empirical volatility are
\begin{equation}
\begin{aligned}
\widehat{\mathrm{RV}}_{i,t,h}
&=\frac{1}{S}\sum_{s=1}^{S}\mathrm{RV}^{(s)}_{i,t,h},\\
\mathrm{Vol}_{i,t,h}
&=10^4\sqrt{\mathrm{RV}_{i,t,h}}.
\end{aligned}
\end{equation}
Empirical variance uses the same grid and log-return definition.
Variance is computed before averaging Monte Carlo samples; it is not
computed from the ensemble mean price path. Reported volatility is
in basis points over the forecast horizon and is not annualized.
Previous-tick sampling for this pathwise statistic differs from the
nearest-event horizon-price rule used for return prediction.

At each horizon, valid prompts are ranked globally by predicted
variance and split into five approximately equal-sized bins.
A fixed identifier order resolves forecast ties. Within each bin,
we report the mean and median observed volatility. Realized zero
variance is retained. A prompt requires defined observed variance
and all 32 generated path variances; undefined grid positions are
not removed or filled from future observations.

The quintile diagnostic evaluates risk ordering; monotone group averages alone do not establish calibration of variance magnitudes.

\paragraph{January--February 2026 evaluation.}
Table~\ref{tab:return_prediction_janfeb} covers 256 stocks, eight
trading days per month, and four forecast origins per day. Panel IC and RankIC are pooled over the stock--origin pairs valid for
all 12 models at each horizon (16,211, 16,209, and 16,201 pairs,
respectively).
M3 Large achieves the highest IC and RankIC at all three horizons.

\begin{table*}[t]
\centering
\small
\setlength{\tabcolsep}{4pt}
\renewcommand{\arraystretch}{1.10}
\definecolor{PredictionBand}{HTML}{F0F2F5}
\definecolor{PredictionOurs}{HTML}{E7EDF6}
\begin{tabular}{p{0.40\textwidth}*{6}{>{\raggedleft\arraybackslash}p{\dimexpr(\textwidth-0.40\textwidth-14\tabcolsep)/6\relax}}}
\toprule
& \multicolumn{2}{c}{\textbf{1 minute}}
& \multicolumn{2}{c}{\textbf{2 minutes}}
& \multicolumn{2}{c}{\textbf{3 minutes}} \\
\cmidrule(lr){2-3}\cmidrule(lr){4-5}\cmidrule(lr){6-7}
Model & IC & RankIC & IC & RankIC & IC & RankIC \\
\midrule
\rowcolor{PredictionBand}
\multicolumn{7}{l}{\footnotesize\bfseries Zero-shot time-series foundation models} \\
\addlinespace[2pt]
Chronos-2$^{\dagger}$ & $0.0203$ & $0.0054$ & $0.0188$ & $-0.0011$ & $0.0113$ & $-0.0167$ \\
TimesFM-2.5$^{\dagger}$ & $0.0060$ & $-0.0169$ & $0.0144$ & $0.0112$ & $0.0227$ & $-0.0090$ \\
Toto-1.0 & $0.0014$ & $0.0152$ & $-0.0032$ & $0.0016$ & $0.0196$ & $0.0060$ \\
\addlinespace[3pt]
\rowcolor{PredictionBand}
\multicolumn{7}{l}{\footnotesize\bfseries Fine-tuned time-series foundation models} \\
\addlinespace[2pt]
Chronos-2$^{\dagger}$ & $0.0359$ & $0.0390$ & $0.0183$ & $0.0290$ & $0.0151$ & $0.0309$ \\
TimesFM-2.5$^{\dagger}$ & $0.0263$ & $0.0203$ & $0.0231$ & $0.0135$ & $0.0200$ & $0.0178$ \\
Toto-1.0 & $-0.0051$ & $-0.0183$ & $0.0073$ & $0.0047$ & $0.0001$ & $-0.0125$ \\
\addlinespace[3pt]
\rowcolor{PredictionBand}
\multicolumn{7}{l}{\footnotesize\bfseries Financial foundation model} \\
\addlinespace[2pt]
Kronos-base & $-0.0172$ & $-0.0177$ & $0.0078$ & $0.0098$ & $-0.0130$ & $-0.0086$ \\
\addlinespace[3pt]
\rowcolor{PredictionBand}\multicolumn{7}{l}{\footnotesize\bfseries Microstructure signals} \\
\addlinespace[2pt]
Weighted-mid displacement & $0.0681$ & $0.0747$ & $\underline{0.0574}$ & $0.0766$ & $\mathbf{0.0467}$ & $0.0631$ \\
Depth-normalized OFI (L1) & $0.0482$ & $0.0622$ & $0.0284$ & $0.0426$ & $\underline{0.0401}$ & $0.0442$ \\
Queue imbalance (L1) & $0.0377$ & $0.0627$ & $0.0429$ & $0.0695$ & $0.0307$ & $0.0531$ \\
\addlinespace[3pt]
\rowcolor{PredictionBand}
\multicolumn{7}{l}{\footnotesize\bfseries Order foundation models} \\
\addlinespace[2pt]
\rowcolor{PredictionOurs}
\textbf{M3 Tiny (ours)} & $\underline{0.0863}$ & $\underline{0.1276}$ & $0.0465$ & $\underline{0.0868}$ & $0.0218$ & $\underline{0.0787}$ \\
\rowcolor{PredictionOurs}
\textbf{M3 Large (ours)}
& $\mathbf{0.1178}$ & $\mathbf{0.1685}$
& $\mathbf{0.0657}$ & $\mathbf{0.1202}$
& $\underline{0.0438}$ & $\mathbf{0.1144}$ \\
\addlinespace[3pt]
\bottomrule
\end{tabular}
\par\vspace{3pt}
\parbox{\textwidth}{\scriptsize\raggedright
Point forecasts use MC32 mean prices. $\dagger$ indicates approximate
quantile sampling.
}
\caption{Return prediction on the January--February 2026 panel.}
\label{tab:return_prediction_janfeb}
\end{table*}

\subsection{Directional Stress Testing}
\label{appx:stress}

We use directional stress testing to evaluate whether the simulator
produces economically consistent price responses under controlled
order-flow imbalance. For each historical prompt, we construct buy-
and sell-stressed continuations from the same initial limit-order-book
state. Under buy stress, the multiplicity of each generated buy-side
add event is set to $k\in\{3,5,10\}$, while sell-side events are left
unchanged. Sell stress is defined symmetrically. Each replicated event
preserves the original price, quantity, and timestamp; cancellation
events are not replicated.

Each replicated event is also encoded back into the autoregressive
context before further generation. The diagnostic therefore captures
both the immediate order-book response and subsequent changes in
generated order sequences.


Let $\widetilde m^c_{i,t}(h)$ denote the displayed mean mid-price
for scenario $c$, and let $m_{i,t}$ denote the initial mid-price.
The corresponding log return is
$
\widetilde r^c_{i,t,h}
=
\log\left(
\frac{\widetilde m^c_{i,t}(h)}{m_{i,t}}
\right).
$
The reported trajectory is the cross-prompt mean of
$10^4\widetilde r^c_{i,t,h}$ in basis points, with 95\% confidence
bands obtained by bootstrapping date--time clusters.
Figure~\ref{fig:whatif_stress} displays the first 30 seconds.

\subsection{TWAP Market Impact}
\label{appx:sqrt}

\paragraph{TWAP protocol.}
For security $i$ on trading day $d$, let $\widehat V_{i,d}$ and
$\widehat\sigma_{i,d}$ denote the separate medians of daily trading
volume and daily realized volatility over the preceding 20 trading
days, excluding the test day. The target meta-order size is
\begin{equation}
\begin{aligned}
Q_{i,t}(\rho)
&=\left\lceil \rho\widehat V_{i,d(t)} \right\rceil,\\
\rho
&\in\{0.001,0.002,\ldots,0.010\}.
\end{aligned}
\end{equation}
Thus, $\rho$ specifies a fraction of estimated daily volume.

Each meta-order is split into 20 child orders submitted at
$9,18,\ldots,180$ seconds. Child orders are priced aggressively
against the current simulated book to ensure that they can be traded, processed by the matching engine,
and encoded back into the autoregressive context. Subsequent background
orders therefore conditions on the intervention history.
After execution ends, simulation continues for another 120 seconds.

\paragraph{Impact measurement.}
Let $\widehat m^{\epsilon,\rho}_{i,t}(h)$ denote the arithmetic mean of
Monte Carlo mid-price predictions under
intervention side $\epsilon\in\{+1,-1\}$,
where $+1$ denotes buy and $-1$ denotes sell.
Each prediction uses the event quote nearest to horizon $h$.
Consistent with Eq.~\eqref{eq:return_forecasts}, direction-adjusted
impact is
\begin{equation}
\begin{aligned}
I^{\epsilon,\rho}_{i,t}(h)
&=
\epsilon\log\left(
\frac{\widehat m^{\epsilon,\rho}_{i,t}(h)}{m_{i,t}}
\right)\\
&=
\epsilon\log\left(
1+\widehat R^{\epsilon,\rho}_{i,t,h}
\right).
\end{aligned}
\end{equation}
Impact is measured relative to the initial mid-price, without
subtracting a baseline return. Monte Carlo prices are averaged before
taking the logarithm. The power-law analysis uses the execution
endpoint $h=180$ seconds.

\paragraph{Normalization and aggregation.}
For each side and size bin, let $\mathcal E_{\epsilon,\rho}$ contain origins
with valid predictions and positive directional impact.
The plotted coordinates are
\begin{equation}
\begin{aligned}
x^\epsilon_\rho
&=
\operatorname{median}_{(i,t)\in\mathcal E_{\epsilon,\rho}}
\left[
\log\left(
\frac{Q_{i,t}(\rho)}{\widehat V_{i,d(t)}}
\right)
\right],\\
y^\epsilon_\rho
&=
\operatorname{median}_{(i,t)\in\mathcal E_{\epsilon,\rho}}
\left[
\log\left(
\frac{I^{\epsilon,\rho}_{i,t}(180)}
     {\widehat\sigma_{i,d(t)}}
\right)
\right].
\end{aligned}
\end{equation}

Normalization and logarithms are applied before aggregation.
Nonpositive impacts are excluded from the logarithmic fit.
The horizontal coordinate uses target submitted quantity.

We fit the ten bin-level points separately for each side using
equal-weight ordinary least squares with a free intercept:
\begin{equation}
y^\epsilon_\rho
=
a_\epsilon+\beta_\epsilon x^\epsilon_\rho
+e^\epsilon_\rho.
\end{equation}
The exponent is estimated freely, with $1/2$ serving as the
square-root reference. Replacing the group medians with means
provides a robustness check.

The accompanying trajectory plots show the median signed normalized
impact $I^{\epsilon,\rho}_{i,t}(h)/\widehat\sigma_{i,d(t)}$, including
nonpositive values, throughout the 300-second simulation.

\end{document}